\documentclass[twocolumn,amsmath,amsfonts,amssymb,aps,prx,preprintnumbers,superscriptaddress]{revtex4-2}

\usepackage[utf8]{inputenc}
\usepackage[T1]{fontenc}
\usepackage{lmodern}
\usepackage{graphicx}
\usepackage{dcolumn}
\usepackage{bm}
\usepackage{textcomp}
\usepackage{float}

\usepackage[normalem]{ulem}
\usepackage{ifpdf}
\usepackage[squaren,Gray]{SIunits}

\usepackage{amsmath}

\ifpdf
\usepackage{epstopdf}
\usepackage[pdftex,unicode,pdfstartview={FitH},pdfborder={0 0 0}]{hyperref}
\usepackage{hypcap}
\else
\usepackage[hypertex]{hyperref}
\fi
\hypersetup{
    bookmarksnumbered = true,
    colorlinks = true, linkcolor = darkblue,
    citecolor = darkblue, filecolor = darkblue,
    menucolor = darkblue, urlcolor = black
}

\usepackage{color}
\definecolor{red}{rgb}{1,0,0}
\definecolor{blue}{rgb}{0,0,1}
\definecolor{black}{rgb}{0,0,0}

\usepackage{color}
\definecolor{darkblue}{rgb}{0,0,0.6}

\definecolor{red}{rgb}{1,0,0}

\definecolor{green}{rgb}{0,0.6,0}

\definecolor{grey}{rgb}{0.7,0.7,0.7}

\definecolor{orange}{rgb}{0.8,0.4,0}

\usepackage{titlesec}           
\titleformat{\subsection}
{\bfseries} 
{}          
{0.0cm}     
{}          
[]          

\begin{document}

\title{Cavity-enhanced photon indistinguishability \\ at room temperature and telecom wavelengths}

\author{Lukas Husel}
\altaffiliation{These authors contributed equally to this work}
\affiliation{Fakult\"at f\"ur Physik, Munich Quantum Center, and Center for NanoScience (CeNS), Ludwig-Maximilians-Universit\"at M\"unchen, Geschwister-Scholl-Platz 1, 80539 M\"unchen, Germany}

\author{Julian Trapp}
\altaffiliation{These authors contributed equally to this work}
\affiliation{Fakult\"at f\"ur Physik, Munich Quantum Center, and Center for NanoScience (CeNS), Ludwig-Maximilians-Universit\"at M\"unchen, Geschwister-Scholl-Platz 1, 80539 M\"unchen, Germany}

\author{Johannes Scherzer}
\affiliation{Fakult\"at f\"ur Physik, Munich Quantum Center, and Center for NanoScience (CeNS), Ludwig-Maximilians-Universit\"at M\"unchen, Geschwister-Scholl-Platz 1, 80539 M\"unchen, Germany}

\author{Xiaojian Wu}
\affiliation{Department of Chemistry and Biochemistry, University of Maryland, Maryland, USA}

\author{Peng Wang}
\affiliation{Department of Chemistry and Biochemistry, University of Maryland, Maryland, USA}

\author{Jacob Fortner}
\affiliation{Department of Chemistry and Biochemistry, University of Maryland, Maryland, USA}

\author{Manuel Nutz}
\affiliation{Qlibri GmbH, Maistr. 67, 80337 M\"unchen, Germany}

\author{Thomas Hümmer}
\affiliation{Qlibri GmbH, Maistr. 67, 80337 M\"unchen, Germany}

\author{Borislav Polovnikov}
\affiliation{Fakult\"at f\"ur Physik, Munich Quantum Center, and Center for NanoScience (CeNS), Ludwig-Maximilians-Universit\"at M\"unchen, Geschwister-Scholl-Platz 1, 80539 M\"unchen, Germany}

\author{Michael F\"org}
\affiliation{Qlibri GmbH, Maistr. 67, 80337 M\"unchen, Germany}

\author{David Hunger}
\email{david.hunger@kit.edu}
\affiliation{Physikalisches Institut, Karlsruhe Institute of Technology, Karlsruhe, Germany}  
\affiliation{Institute for Quantum Materials and Technologies (IQMT), Karlsruhe Institute of Technology (KIT),  Herrmann-von-Helmholtz Platz 1, 76344 Eggenstein-Leopoldshafen, Germany}

\author{YuHuang Wang}
\email{yhw@umd.edu}
\affiliation{Department of Chemistry and Biochemistry, University of Maryland, Maryland, USA}

\author{Alexander H\"ogele}
\email{alexander.hoegele@lmu.de}
\affiliation{Fakult\"at f\"ur Physik, Munich Quantum Center, and Center for NanoScience (CeNS), Ludwig-Maximilians-Universit\"at M\"unchen, Geschwister-Scholl-Platz 1, 80539 M\"unchen, Germany}
\affiliation{Munich Center for Quantum Science and Technology (MCQST), Schellingstr. 4, 80799 M\"unchen, Germany}

\begin{abstract}
Indistinguishable single photons in the telecom-bandwidth of optical fibers are indispensable for long-distance quantum communication. Solid-state single photon emitters have achieved excellent performance in key benchmarks, however, the demonstration of indistinguishability at room-temperature remains a major challenge. Here, we report room-temperature photon indistinguishability at telecom wavelengths from individual nanotube defects in a fiber-based microcavity operated in the regime of incoherent good cavity-coupling. The efficiency of the coupled system outperforms spectral or temporal filtering, and the photon indistinguishability is increased by more than two orders of magnitude compared to the free-space limit. Our results highlight a promising strategy to attain optimized non-classical light sources.
\end{abstract}

\maketitle

\section*{INTRODUCTION}
The capability of two indistinguishable single photons to interfere on a balanced beam splitter and exit jointly on either one of its output ports is a premise to  
quantum photonic applications~\cite{Bouchard2020} such as quantum teleportation~\cite{Bennett1993}, quantum computation~\cite{Obrien2007} or quantum optical metrology~\cite{Dowling2008}. Solid-state based sources of indistinguishable single photons have witnessed tremendous progress in the past decades~\cite{Aharonovic2016}, and among them semiconductor quantum dots stand out as they enable the generation of pure and indistinguishable single photons~\cite{Senellart2017,Arakawa2020} when coupled to optical microcavities~\cite{Somaschi2016,Wang2019,Tomm2021}. However, their operation is so far restricted to cryogenic temperatures and wavelengths in the near-infrared. These limitations motivate alternative platforms operating at ambient conditions and telecom wavelengths to facilitate long-distance quantum communication in optical fibers at reduced loss. Various quantum emitters have proven capable of emitting pure telecom-band single photons at room temperature, including color centers in silicon carbide \cite{Wang2018} and gallium nitride~\cite{Zhou2018}. Recently, the realm of such emitters has been expanded by luminescent nanotube defects (NTDs) in sp$^3$-functionalized single-wall carbon nanotubes~\cite{He2018}. Unlike other emitters, NTDs allow for precise control over the emission wavelength via covalent side-wall chemistry~\cite{He2017,Settele2021,Kwon2016}. Moreover, carbon nanotubes are straightforward to integrate with gated structures~\cite{Li2022}, microcavities~\cite{Pyatkov2016,Jeantet2016,Huemmer2016,Ishii2018} or plasmonic cavities~\cite{Luo2019}. These properties, combined with high single photon purity~\cite{He2017,Luo2019}, render NTDs excellent candidates for the development of sources of quantum light.

As common to solid-state quantum emitters, NTDs are subject to strong dephasing at room temperature. As a result, the coherence time $T_2$ of the emitted photons is orders of magnitude smaller than the population lifetime $T_1$. The respective photon indistinguishability, which can be quantified by $T_2/(2T_1)$~\cite{Bylander2003, Sun2009}, is therefore limited to vanishingly small values. This limitation represents a major challenge in the development of single photon sources based on NTDs and other solid-state quantum emitters. The strategy of reducing $T_1$ to enhance the photon indistinguishability via Purcell enhancement~\cite{Senellart2017} has been successfully applied to quantum dots and Erbium ions in various cavity geometries~\cite{Somaschi2016, Liu2018, Wang2019, Dusanowski2020, Tomm2021, Lakshminarayan2021, Ourari2023} as well as to NTDs by coupling to a plasmonic nanocavity~\cite{Luo2019}. However, all these experiments were operated in the regimes of coherent or incoherent bad cavity coupling~\cite{Auffeves2010}, where strong dephasing at ambient conditions limits both photon coherence time and Purcell enhancement, and thus all experiments to date crucially relied on operation at cryogenic temperatures with reduced dephasing. Although at ambient conditions spectral or temporal filtering of mainly incoherent photons would increase the photon coherence in principle, it would come at the cost of drastically reduced collection efficiency. Therefore, enhancement of $T_2$ at efficiencies exceeding those attainable through spectral or temporal filtering has remained elusive for quantum emitters subject to strong dephasing.

Here, we demonstrate enhancement of photon indistinguishability for telecom-band single photons from individual NTDs coupled to an optical microcavity. Motivated by a recent theoretical proposal, we operate the NTD-cavity system in the regime of incoherent good cavity coupling~\cite{Grange2015}, where the photon coherence time is determined by the cavity linewidth. By choosing a cavity with a spectrally narrow linewidth, we enhance $T_2$ and thus the photon indistinguishability of the coupled NTD-cavity system. At the same time, the cavity enhances the emission via the Purcell effect, thus yielding simultaneous increase of both indistinguishability and efficiency unattainable by spectral or temporal filtering. As a consequence, the efficiency of our system outperforms spectral or temporal filtering within the same bandwidth by at least a factor of four, with an estimated increase of photon indistinguishability by two orders of magnitude as compared to free-space NTDs. Our results experimentally establish the regime of incoherent good cavity-coupling as a powerful strategy for optimized sources of quantum light.

\section*{RESULTS}

\begin{figure*}[t]
	\centering
	\includegraphics[scale=1]{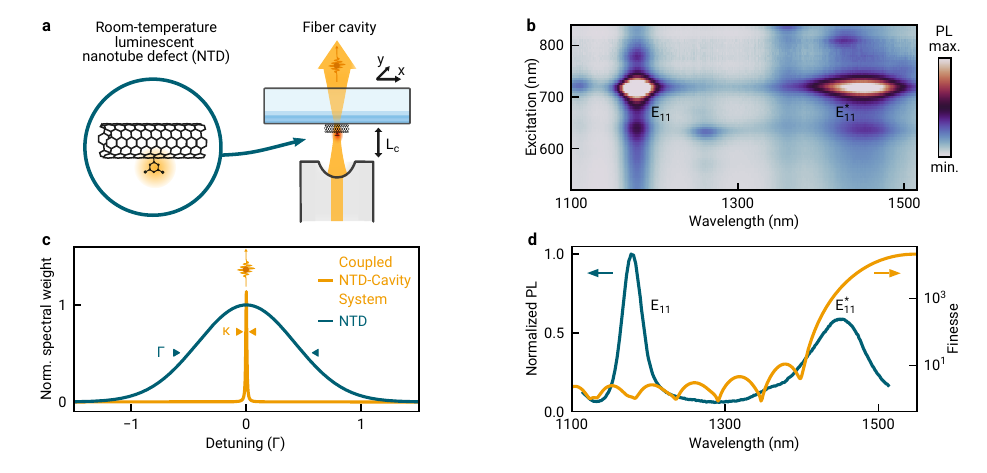}
	\caption{\textbf{Functionalized carbon nanotubes in an open micro-cavity.} \textbf{a}, Schematic of luminescent nanotube defects (NTDs) coupled to the fiber-based open micro-cavity system with tunable cavity length $L_c$ and lateral displacement degrees of freedom of the macroscopic mirror $x$ and $y$. \textbf{b}, Photoluminescence (PL) excitation of functionalized (8,6)~carbon nanotubes with emission band of fundamental excitons ($\mathrm{E_{11}}$) and NTD states ($\mathrm{E_{11}^{*}}$).  \textbf{c}, Schematic spectral weight of strongly dephased free-space NTD luminescence (dark green) subjected to incoherent cavity coupling (orange). \textbf{d}, Ensemble PL spectrum (dark green) and cavity finesse in transfer-matrix simulations (orange). The NTD luminescence spectrally close to maximal cavity finesse was excited at the $E_{11}$ transition at near-unity transmission of the cavity mirrors.}
	\label{fig1}
\end{figure*}

The NTDs used in this work, shown schematically in the left panel of Fig.~\ref{fig1}a, were obtained by functionalizing (8,6)~carbon nanotubes by diazonium reaction \cite{piao_brightening_2013, Wang2022} (see the Methods section for details). The photoluminescence (PL) excitation map of an aqueous suspension with covalently functionalized carbon nanotubes is shown in Fig.~\ref{fig1}b, with an excitation resonance at $718$~nm corresponding to the $E_{\mathrm{22}}$ transition and emission via $E_{\mathrm{11}}$ around $1170$~nm, characteristic of (8,6)~chiral tubes~\cite{Weisman2003}. The red-shifted emission peak, labelled as $E_{11}^*$ and centered at $1470$~nm, corresponds to the luminescence from excitons localized at nanotube side-wall defects with emission wavelength tuned to the telecom \mbox{S-band}~\cite{ITU2016} by the choice of the functional group, in this case the 3,4,5-trifluoro-2-chlorosulfonyl-aryl group paired with the hydroxy group \cite{Wang2022}. For integration in a fiber-based Fabry-Pérot cavity \cite{Hunger2010} shown schematically in the right panel of Fig.~\ref{fig1}a, the nanotubes were dispersed onto a planar macroscopic mirror with a polystyrene layer on top (see the Methods section for details) to ensure optimal coupling near the antinode of the intra-cavity field. Both spectral and spatial overlap between individual NTDs and the fundamental Gaussian cavity mode were optimized by lateral displacement of the macro-mirror and vertical tuning of the fiber-based micro-mirror via piezoelectric actuators. Photons emitted by the NTD-cavity system were coupled into a single mode fiber upon transmission through the planar mirror.

To implement the regime of incoherent NTD-cavity coupling, we employed a distributed Bragg reflector (DBR) mirror coating for spectrally narrow cavity linewidth at the target wavelength of telecom-band emission. Fig.~\ref{fig1}d shows jointly the ensemble PL spectrum and the cavity finesse obtained from a transfer matrix simulation of the DBR coating. In the cavity, the NTD states were excited resonantly through the $E_{11}$ transition at near-unity DBR mirror transmission and thus independent of the cavity resonance condition. With finesse values on the order of $1000$ at the $E_{11}^*$ transition wavelength, the cavity mode provided the primary radiative decay channel for the NTD emission. A combination of long-pass filters was used to suppress the excitation laser and other emission at wavelengths below~\mbox{1400 nm} before detection. 

The effect of cavity-coupling on the photonic spectral bandwidth is illustrated in Fig.~\ref{fig1}c. At ambient conditions, the spectral width of the NTD emission profile is dominated by pure dephasing at rate $\gamma^*$, with ${\Gamma = 2\gamma^*}$ on the order of ten nanometers or $10$~meV. This is orders of magnitude larger than the experimental cavity linewidth, which was determined as ${\kappa = 35.4 \pm 0.1 \,\mu}$eV for the lowest accessible longitudinal mode order, corresponding to ${61.7 \pm 0.2}$~pm in the wavelength domain. The small value of $\kappa$ enables operation of our system in the regime of incoherent good cavity coupling, where ${2g \ll \gamma + \gamma^* + \kappa}$ and ${\kappa < \gamma + \gamma^*}$ holds for the light-matter coupling strength $g$, the population decay rate $\gamma$, $\kappa$ and $\gamma^*$ (see Supplementary Note 1 for details). In this regime, the cavity is incoherently pumped upon initial (incoherent) excitation of the NTD at a rate $R \approx 4g^2/\gamma^*$~\cite{Grange2015}, which in our system is smaller than the population decay rate. Any photon that is coupled into the resonator will be emitted via the cavity mode on a timescale 1/$\kappa$. Since the emission process from the cavity is coherent~\cite{Grange2015}, this constitutes a giant increase in the photon coherence time compared to the free-space limit of $1/\gamma^*$. In the spectral domain, the effect corresponds to a drastic spectral purification as illustrated in Fig.~\ref{fig1}c, similar to spectral filtering. This effect is a key feature of the incoherent good cavity coupling regime and is instrumental for enhanced photon indistinguishability.

In addition to the coherence time, the cavity also enhances the emission spectral density, with the enhancement quantified by the Purcell factor $F_p \propto g^2$ \cite{Kaupp2013} (see Supplementary Note 3 for details). Increasing $F_p$ via the light-matter coupling strength $g$ increases the single photon efficiency, i.e. the probability that a photon is emitted into the cavity mode. In the incoherent good cavity regime, this probability is smaller than the free-space quantum yield due to the large mismatch in the spectral bandwidths of the emitter and the cavity. However, as we demonstrate in the following, maximizing $g$ (achieved in our case by minimizing the microcavity mode volume) results in an efficiency which by far exceeds that obtained by filtering at a spectral bandwidth $\kappa$ or an equivalent temporal bandwidth.

\begin{figure}[t]
	\centering
	\includegraphics[scale=1]{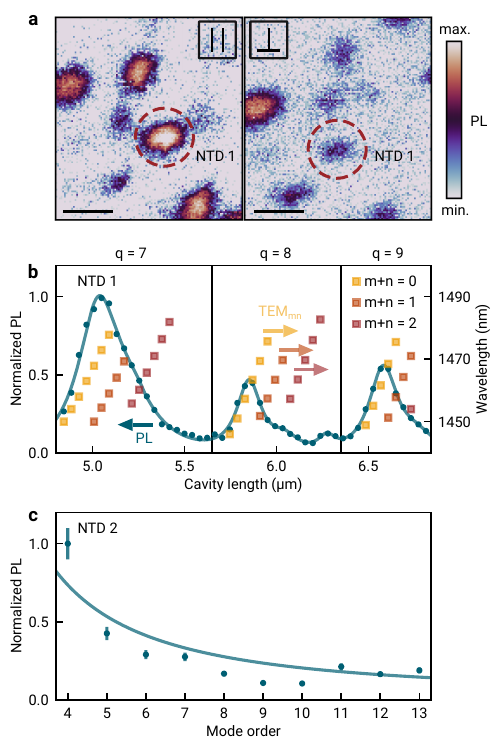}
	\caption{\textbf{Photoluminescence characteristics of cavity-coupled carbon nanotube defects.} \textbf{a}, Cavity-enhanced PL raster-scan maps recorded for two orthogonal linear polarizations. The detection basis is chosen parallel (left) and perpendicular (right) to the axis of the emitter NTD~1 marked by the dashed circle. The scalebar is $5 \mu$m \textbf{b}, Normalized PL of NTD~1 as a function of the cavity length, tuned over three longitudinal mode orders (blue circles). The emission spectrum is probed at the resonance wavelengths of the transverse electromagnetic (TEM) cavity modes (yellow, orange and red squares). The solid line was obtained from the fit described in the main text. The colored arrows indicate the respective y-axis \textbf{c}, Maximum PL intensity of a different emitter (NTD~2) as a function of the longitudinal mode order, normalized by the coupling efficiency into the single mode fiber. Cavity-enhancement of the PL intensity is inversely proportional to the mode volume $V_{\mathrm{c}}$, as evident from best-fit (solid line). The error bars give the standard uncertainty, dominated by experimental uncertainty in fiber coupling.}
	\label{fig2}
\end{figure}

Individual NTDs were identified in the cavity from maps of PL intensity as in Fig.~\ref{fig2}a, recorded upon lateral raster-scan displacement of the macroscopic mirror for a fixed cavity length. The two maps of Fig.~\ref{fig2}a were acquired for two orthogonal linear polarizations in the detection path and feature bright PL spots with lateral extent given by the point spread function of the Gaussian fundamental cavity mode with a waist of 2~$\mu$m. The left (right) map in Fig.~\ref{fig2}a was obtained for parallel (orthogonal) orientation of the polarization axis with respect to the nanotube with NTD~1. The contrast in the brightness between the two maps for most PL hot-spots indicates a large degree of linear polarization at the emission sites, a hallmark of the well-known antenna effect in individual carbon nanotubes~\cite{Hoegele_photon_2008, Hartschuh2003}. 

In Fig.~\ref{fig2}b, we show the normalized PL intensity of NTD~1 as the cavity length is tuned over three longitudinal mode orders $q = 7$, $8$ and $9$. For each mode order, we observe an asymmetric emission profile, stemming from higher order transverse electromagnetic (TEM) modes. Since the cavity linewidth $\kappa$ is much smaller than the emitter PL linewidth $\Gamma$, the NTD emission spectrum is probed at the resonance wavelength of each TEM-mode~\cite{Auffeves2009} with resonance wavelengths given explicitly on the right axis of Fig.~\ref{fig2}b. We fitted the data by the sum of three Lorentzians for each longitudinal mode order, with the result shown as the solid line in Fig.~\ref{fig2}b (TEM$_{mn}$ mode orders with $n+m > 2$ were neglected due to vanishing contributions). From the fit, we obtained the emission wavelength $1465 \pm 3$~nm, and a FWHM linewidth of $28 \pm 5$~nm, corresponding to $\gamma^* = 8 \pm 2$~meV.

Fig.~\ref{fig2}c shows the effect of the cavity mode volume on the photon emission efficiency. We measured the collected PL intensity of a different NTD with comparable brightness for ten consecutive longitudinal mode orders, normalized to the largest value and corrected for the variation of the measured fiber coupling. The fiber coupling efficiency depends on the mode waist, which in turn changes with cavity length. We observed an increase in the PL intensity by a factor of six as the cavity was tuned to the  lowest accessible longitudinal mode order $q = 4$. This mode order corresponds to an intermirror separation of $2.6~\mu$m, mainly limited by the profile depth of the fiber mirror of $2~\mu$m. The increase in the PL intensity stems from an enhancement in light-matter coupling strength $g$ as the cavity length and hence the mode volume $V_{\mathrm{c}}$ is decreased. For our regime of low Purcell enhancement, where the NTD population lifetime is mainly unaffected by the cavity, the emission intensity is proportional to $g^2$, which in turn is inversely proportional to $V_{\mathrm{c}}$ (see Supplementary Note 3 for details). A fit of $\alpha V_{\mathrm{c}}^{-1}$, with $V_{\mathrm{c}}$ calculated from the cavity length $L_{\mathrm{c}} = q\lambda/2$~\cite{Hunger2010} and the amplitude $\alpha$ as a free fit parameter, yields a good correspondence with the data (solid line in Fig.~\ref{fig2}c).

\begin{figure}[t]
	\centering
	\includegraphics[scale=1]{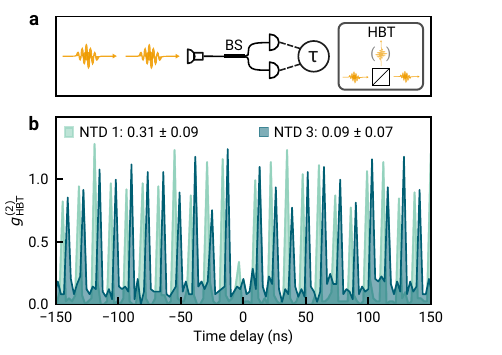}
	\caption{\textbf{Telecom-band room-temperature single photons from the coupled NTD-cavity system.} \textbf{a} Schematic of a Hanbury-Brown-Twiss (HBT) setup based on a fiber beamsplitter (BS). \textbf{b}, HBT autocorrelation function of cavity-coupled NTD~1 (light green) and NTD~3 (dark green), with second order coherence at zero time delay $g^{(2)}_{\mathrm{HBT}}(0) = 0.31 \pm 0.09$ and $0.09 \pm 0.07$, respectively.}
	\label{fig3}
\end{figure}

Operating the coupled NTD-cavity system at maximum cavity-enhancement of the PL intensity, we determined second-order correlations in photon emission events with a fiber-based Hanbury-Brown-Twiss (HBT) setup shown schematically in Fig.~\ref{fig3}a. Photons generated via pulsed laser excitation were coupled into a fiber beamsplitter, and detection events at the output ports were time-correlated to obtain the normalized second-order autocorrelation function $g^{(2)}_{\mathrm{HBT}}(\tau)$. The shot-noise limited results of the HBT experiment on two distinct NTDs are shown in Fig.~\ref{fig3}b, with the corresponding antibunching values $g^{(2)}_{\mathrm{HBT}}(0) = 0.31 \pm 0.09$ and $0.09 \pm 0.07$ as measures of the single photon purity.

\begin{figure}[H]
	\centering
	\includegraphics[scale=1]{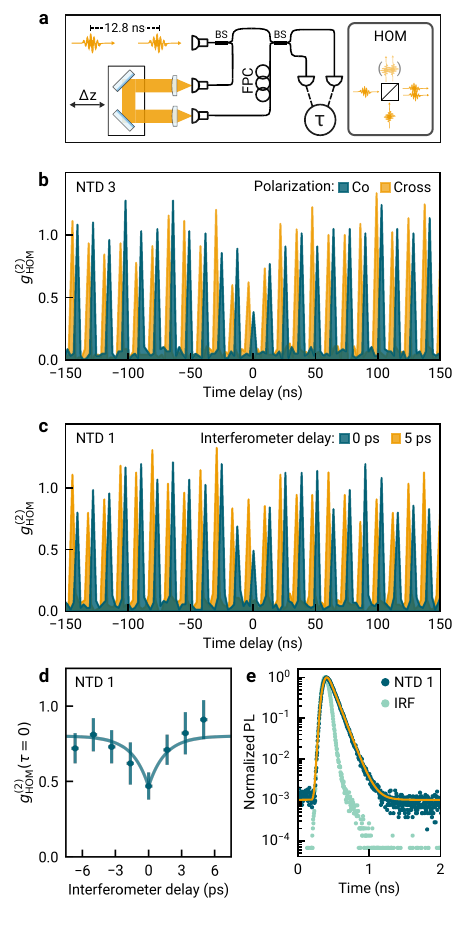} 
	\label{fig4_0}	
	\vspace*{-7mm}
	\caption{\textbf{Demonstration of cavity-enhanced photon indistinguishability.} \textbf{a}, Schematic of the imbalanced Mach-Zehnder interferometer to probe the photon indistinguishability in Hong-Ou-Mandel (HOM) type experiments based on fiber beamsplitters (BS). The time delay between the interferometer arms was tuned via the displacement $\Delta z$, and their relative polarization by the fiber polarization controller (FPC) in one arm. \textbf{b}, HOM autocorrelation function of NTD~3 for co-polarized (dark green) and cross-polarized (orange) interferometer arms with delay of one excitation pulse. The difference in the coincidence probabilities at zero-delay is a hallmark of two-photon interference with visibility ${v = 0.51 \pm 0.21}$. \textbf{c}, HOM autocorrelation function of NTD~1, measured in co-polarized interferometer configuration for interferometer delays 0~ps (dark green) and 5~ps (orange). Zero interferometer delay again corresponds to delay by one excitation pulse separation. \textbf{d}, HOM autocorrelation function at time delay $\tau = 0$ for NTD~1 as a function of the interferometer delay, with visibility ${v = 0.65 \pm 0.24}$. The solid line is an empirical fit to the HOM dip described in the main text. The horizontal error bars correspond to the standard uncertainty in the interferometer delay; the vertical error bars correspond to the standard uncertainty determined as described in the Methods section. \textbf{e}, Temporal PL decay of NTD~1 (dark green data) and instrument response (light green data). The orange line shows the result of a biexponential decay model.}
	\label{fig4}
\end{figure}

The photon indistinguishability was quantified in Houng-Ou-Mandel (HOM) type experiments using an imbalanced Mach-Zehnder interferometer shown schematically in Fig.~\ref{fig4}a. The train of single photon pulses generated by the source was first split in a fiber beamsplitter. The time delay $\Delta t$ in the interferometer was tuned by the path difference $\Delta z$ with an adjustable delay stage to enable two-photon interference between consecutively emitted photons at the second beam splitter. In this setting, a delay of zero implies a separation by one excitation pulse. The relative polarization between the interferometer arms was set by fiber polarization controllers, and the detection events at the output ports were time-correlated to obtain the HOM-autocorrelation function $g^{(2)}_{\mathrm{HOM}}(\tau)$ (see the Methods section for details). 
First, we initialized the interferometer at zero delay and performed a two-photon interference experiment for co- and cross-polarized interferometer arms on NTD~3. The shot-noise limited results are shown in Fig.~\ref{fig4}b. For the co-polarized configuration, we observe a reduction of the measured correlations at zero time delay. This is a hallmark of quantum coherent two-photon interference: the (partially) indistinguishable single photons arriving simultaneously at different input ports of the beamsplitter are likely to exit at the same output port, resulting in reduced correlations at zero time delay \cite{Santori2002,Somaschi2016,Wang2019}. We quantify the respective degree of the photon indistinguishability by the two-photon interference visibility $v$ that one would detect in an interferometer with balanced beamsplitters and unity classical visibility~\cite{Tomm2021}. We obtain $v = 0.51 \pm 0.21$ for the data in Fig.~\ref{fig4}b, taking into account non-identical reflection and transmission of the beamsplitters and finite single photon purity of NTD~3 (see Supplementary Note 4 for details). 

Successively, we performed the HOM interference experiment for varying interferometer delays on NTD~1, with autocorrelation histograms for interferometer delays of 0 and 5~ps shown in Fig.~\ref{fig4}c. The observed reduction in correlations at zero time delay is again a hallmark of two-photon interference, where tuning between the two interferometer delay settings is approximately equivalent to switching the polarization configuration as in Fig.~\ref{fig4}b. In Fig.~\ref{fig4}d, we show the measured value of the HOM autocorrelation function at zero time delay for varying interferometer delay. Upon transition through zero-delay, we observed the characteristic HOM dip due to reduced cross-channel correlations by two-photon interference, described by the empirical formula ${c[1-a\exp(-|\Delta t|/\tau_{\mathrm{HOM}})]}$, where $a$ is an amplitude, $c$ is an offset at large interferometer delays $\Delta t$, and $\tau_{\mathrm{HOM}}$ is the characteristic timescale of the HOM interference \cite{Santori2002}. From the best fit to the data shown by the solid line in Fig.~\ref{fig4}d, we determined $\tau_{\mathrm{HOM}} = 2 \pm 2$~ps, and a visibility of $0.65 \pm 0.24$ (see Supplementary Note 4 for details), consistent with the value of $0.51 \pm 0.21$ for NTD~3. 

The characteristic two-photon interference time scale $\tau_{\mathrm{HOM}}$ is given by the jitter in the photon arrival time at the beamsplitter, which in turn is determined by the population lifetime~\cite{Bylander2003} (see Supplementary Note 5 for details). For the emitter NTD~1, the fit to the data in Fig.~\ref{fig4}d thus implies a population decay within a few picoseconds. This time scale can be associated with the short decay component of the biexponential PL decay characteristic for NTDs~\cite{He2017,Hartmann2016}. The fast and slow decay channels with time constants $\tau_{\mathrm{fast}}$ and $\tau_{\mathrm{slow}}$ arise from an interplay of bright and dark exciton reservoirs, with $\tau_{\mathrm{fast}}$ as short as a few picoseconds and relative decay amplitudes close to unity in larger-diameter nanotubes \cite{Hartmann2016}. In Fig.~\ref{fig4}e, we show by the solid line the result of a cavity-coupled biexponential model decay with $\tau_{\mathrm{fast}} = 2$~ps and $\tau_{\mathrm{slow}} = 91$~ps, convoluted with the instrument response function, together with the measured PL decay for NTD~1.

Although the short decay component is not resolved directly in the instrument-response limited data of Fig.~\ref{fig4}e, the identification of $\tau_{\mathrm{fast}}$ with $\tau_{\mathrm{HOM}}$ is plausible. In the framework of the incoherent good cavity regime, the feeding of the cavity through the fast decay channel generates photons with near-unity visibility~\cite{Grange2015}. The actual visibility in Fig.~\ref{fig4}d is lower than unity ($0.65 \pm 0.24$), most probably due to photons generated via the slow process with lifetimes exceeding the cavity coherence time of $20$~ps, which renders them partly distinguishable. A reduction in visibility is also backed by our model for time-dependent NTD-cavity coupling, which predicts $v=0.3$ for the NTD~1 in Fig.~\ref{fig4}d (see Supplementary Note 4 for details). The deviation between measured and estimated value is consistent with operation of our experiment at wavelengths on the edge of the DBR stopband (see Fig.~\ref{fig1}c). In this regime, small shifts towards larger resonance wavelength caused by cavity length drifts can decrease the cavity linewidth by a factor of up to two and in turn result in increased visibility, which is inversely proportional to $\kappa$~\cite{Grange2015}.

The visibility in the two-photon interference data in Fig.~\ref{fig4}d corresponds to a $217$-fold enhancement of the value estimated for the free-space limit (see Supplementary Note 4 for details). For spectrally filtered free-space emission, the same visibility can be achieved in principle, yet at the cost of very low single photon efficiency. In the incoherent good cavity regime implemented here, the measured lower bound $\mathrm{min}(\beta_{\mathrm{c}}) = (4.0 \pm 0.1) \cdot 10^{-3}$ and expected value $\beta_{\mathrm{c}} = 6.6 \cdot 10^{-3}$ for the Purcell-enhanced single photon efficiency are a factor of four and seven larger than the estimated upper bound ${\beta_{\mathrm{fs}} = \kappa/(\pi\gamma^*) = 1\cdot 10^{-3}}$ for spectrally filtered free-space decay, whose actual value we expect to be at least one order of magnitude smaller when taking into account the non-unity NTD quantum yield (see Supplementary Note 3 for details). Further benefit arises from the fiber-based design of our cavity, which in principle allows unity in-fiber coupling efficiency in contrast to free-space collection with inherent diffraction losses. 

\section*{DISCUSSION}

To conclude, we have presented a room-temperature source of telecom-band single photons with emission efficiency and indistinguishability drastically enhanced by incoherent NTD-cavity coupling. To our knowledge, our results represent the first demonstration of cavity-enhanced indistinguishability for a quantum emitter with room-temperature dephasing. 
We estimate that the current two-photon interference visibility of about $0.5$ can be improved to near-unity values by increasing the cavity finesse to $35000$, a feasible value with open fiber cavites~\cite{Huemmer2016}. Simultaneously, a further reduction of the mode volume to recently reported values~\cite{Kaupp2016} would yield an enhancement in emission efficiency by another order of magnitude. Even without these improvements, our results represent a major step towards room-temperature quantum photonic devices for applications at telecom-wavelength in optical quantum computation~\cite{Knill2001} or long-distance communication relying on optical quantum repeaters~\cite{Azuma2015}.

\vspace{14pt}
\noindent \textbf{METHODS}\\

\noindent \textbf{Sample preparation}

\noindent The NTDs were prepared by functionalizing (8,6) carbon nanotubes based on a method we reported previously \cite{Wang2022}. Briefly, raw HiPco SWCNT material (NoPo Nanotechnologies, India) was dissolved in chlorosulfonic acid (99\%, Sigma-Aldrich) at a concentration of 0.5 mg/mL, followed by adding 2-amino-4,5,6-trifluorobenezen-1-sulfonyl chloride, which was synthesized from 3,4,5-trifluoeoaniline, and NaNO2 (ReagentPlus® > 99.0\%, Sigma-Aldrich) to concentrations of 0.24 mg/mL and ~0.2 mg/mL, respectively. After fully mixed, the acid mixture was then added drop-by-drop to Nanopure® water with vigorous stirring, resulting in the formation of NTD functionalized carbon nanotubes that precipitated from the solution as black precipitates. The precipitates were filtered and rinsed with an excessive amount of Nanopure® water. The synthesized NTDs were dissolved in 2\% (wt/v) sodium deoxycholate (DOC, Sigma-Aldrich, $\geq$ 97\%) solution and centrifuged at 16400 rpm for 1 h to remove any bundles. The nanotubes with NTDs were then sorted by aqueous two-phase extraction \cite{Wang2022, Zheng2015} in a solution of 2\% (w/v) DOC in deuterium oxide (D2O, Cambridge Isotope Laboratories, Inc. 99.8\%) to obtain NTDs on (8,6) chirality enriched nanotubes. 

\noindent Next, a macroscopic planar mirror was spin-coated with a 10 $\mu$L solution of 3\% (wt/v) polystyrene/toluene, at 2000 RPM for 1 minute, resulting in the formation of a polystyrene spacer layer estimated to be ~150 nm thick. The coated mirror was then vacuum-dried at room temperature for 24 hours before being deposited with 5 $\mu$L of the NTDs containing solution by spin-coating at 3000 RPM for 1 minute. \\

\noindent \textbf{Fiber-based cavity} 

\noindent The experiments were conducted in an ultra-stable fiber-based open-cavity platform (\textit{Qlibri Quantum}, Qlibri GmbH). The cavity is formed by a microscopic concave fiber mirror with a radius of curvature of $25~\mu$m, fabricated by CO$_\mathrm{2}$ laser ablation \cite{Hunger2010, Hunger2012}, and a macroscopic planar mirror with a $150$~nm thick polystyrene spacer layer and functionalized carbon nanotubes on top. The spacer layer was included to place NTDs close to an antinode of the intra-cavity field. Three translational degrees of freedom are accessible through piezoelectric positioners, allowing for lateral scans and length-tuning of the cavity with sub-nanometer precision. Fiber and sample mirror have identical DBR coatings, designed for high reflectivity at telecom wavelengths (minimum transmission $T=95.2$~ppm at wavelength of $1535.4$~nm) and fabricated by ion beam sputtering (Laseroptik GmbH). At a wavelength of $1468$~nm, close to the $E_{11}^{*}$ peak maximum, the largest measured finesse was $3010 \pm 10$ for the lowest accessible longitudinal mode order $q = 4$. For this mode order, corresponding to a mirror distance of $L_{\mathrm{c}} = 2.6~\mu$m, we calculated a mode waist of $\omega_0 = 2~\mu$m and a cavity mode volume of $V_{\mathrm{c}} = 8.2~\mu \mathrm{m}^3$~\cite{Hunger2010}.\\

\noindent \textbf{Photoluminescence and photon correlation experiments}

\noindent PL measurements were performed under resonant excitation of the $E_{11}$ transition using a pulsed supercontinuum white light source (NKT SuperK Extreme) at a repetition rate of 78 MHz that was spectrally filtered in a home-built monochromator to a linewidth of 2 nm. The cavity was tuned on resonance with the $E_{11}^{*}$ transition by changing the mirror distance. 
The PL emitted through the planar mirror of the cavity was collimated by an achromatic doublet lens (Thorlabs AC127-019-C-ML), filtered with two longpass filters (Thorlabs FEL1400, band edge 1400~nm, and Semrock BLP02-1319R-25, band edge 1320~nm) and coupled into a single mode fiber. Detection was performed with a pair of superconducting nanowire single photon detectors (Scontel TCOPRS-CCR-SW-85) and time-correlated with a TCSPC module (Swabian Instruments Time Tagger Ultra and PicoQuant PicoHarp300). Second-order photon correlation measurements were performed in a standard Hanbury-Brown-Twiss configuration. For Hong-Ou-Mandel type two-photon interference experiments, a home-built fiber-based imbalanced Mach-Zehnder interferometer was employed. A mechanical delay stage was used to tune the interferometer delay on sub-picosecond scale. Polarization was set by fiber-polarization controllers (Thorlabs FPC562). 

Photon correlation histograms were obtained by integrating detection events in 2.5 ns wide windows. The resulting histograms feature prominent peaks separated by the delay between the excitation pulses. To obtain the correlation functions $g^{(2)}_{\mathrm{HBT}}$ and $g^{(2)}_{\mathrm{HOM}}$, we normalized the histograms with respect to the average height of histogram peaks $N_{\mathrm{\infty}}$ at large time delays. The standard uncertainty of the measured peak height $N_0$ at $\tau =0$ is given by $\sqrt{N_0}$ \cite{Fischer2016} and is the dominant uncertainty in the measurement of $N_0$. The standard uncertainty in quantities derived from measured peak heights was obtained by Gaussian error propagation, considering the uncertainties in all input parameters. The normalized second-order correlation at zero time delay $g^{(2)}(0)$ was obtained from the measured histograms as $g^{(2)}(0) = N_0/N_{\mathrm{\infty}}(1\pm 1/\sqrt{N_0})$ \cite{Fischer2016} including dark count and background correction~\cite{Somaschi2016}. The uncertainties in $N_{\mathrm{\infty}}$ and background were found to have negligible influence on this measurement, whose uncertainty is dominated by the uncertainty in $N_0$.  \\

\noindent \textbf{ACKNOWLEDGEMENTS} 

\noindent We gratefully acknowledge helpful discussions with Lukas Knips and support by Max Huber for manufacturing of the cavity. This research was funded by the European Research Council (ERC) under the Grant Agreement No.~772195 as well as the Deutsche Forschungsgemeinschaft (DFG, German Research Foundation) within the Germany's Excellence Strategy EXC-2111-390814868. L.\,H. and A.\,H. acknowledge funding by the Bavarian Hightech Agenda within the EQAP project. B.\,P. acknowledges funding by IMPRS-QST. D.H. acknowledges support by the Karlsruhe School of Optics \& Photonics (KSOP). Y.W. gratefully acknowledges the U.S. National Science Foundation for funding support (grant no. PHY1839165 and CHE2204202).\\

\noindent \textbf{AUTHOR CONTRIBUTIONS}

\noindent D.~H., Y.~W. and A.~H. conceived the project. L.~H. and J.~T. set up and performed the experiments with contributions by J.~S., evaluated the data, and carried out theoretical analysis and modelling. X.~W. led the sample preparation of defect-tailored carbon nanotubes synthesized by P.~W. with contributions from J.~F. and supervision by Y.~W.. B.~P. contributed to the initial sample characterization by optical spectroscopy. M.~N., T.~H. and M.~F. designed and manufactured the cavity and provided support for its operation. L.~H., J.~T., D.~H. and A.~H. analyzed the data. L.~H., J.~T. and A.~H. wrote the manuscript. All authors commented on the manuscript. \\

\noindent \textbf{DATA AVAILABILITY}

\noindent The source data generated in this study have been deposited in the LMU Open Data database under accession code doi.org/10.5282/ubm/data.460. \\

\noindent \textbf{CODE AVAILABILITY}

\noindent The codes that support the findings of this study are available from the corresponding authors upon reasonable request.\\

\noindent \textbf{COMPETING INTERESTS}

\noindent The authors declare no competing interests. \\


\begin{thebibliography}{10}
    \expandafter\ifx\csname url\endcsname\relax
    \def\url#1{\texttt{#1}}\fi
    \expandafter\ifx\csname urlprefix\endcsname\relax\def\urlprefix{URL }\fi
    \providecommand{\bibinfo}[2]{#2}
    \providecommand{\eprint}[2][]{\url{#2}}
    
    \bibitem{Bouchard2020}
    \bibinfo{author}{Bouchard, F.} \emph{et~al.}
    \newblock \bibinfo{title}{Two-photon interference: the {Hong}–{Ou}–{Mandel}
        effect}.
    \newblock \emph{\bibinfo{journal}{Reports on Progress in Physics}}
    \textbf{\bibinfo{volume}{84}}, \bibinfo{pages}{012402}
    (\bibinfo{year}{2020}).
    
    \bibitem{Bennett1993}
    \bibinfo{author}{Bennett, C.~H.} \emph{et~al.}
    \newblock \bibinfo{title}{Teleporting an unknown quantum state via dual
        classical and {Einstein}-{Podolsky}-{Rosen} channels}.
    \newblock \emph{\bibinfo{journal}{Physical Review Letters}}
    \textbf{\bibinfo{volume}{70}}, \bibinfo{pages}{1895--1899}
    (\bibinfo{year}{1993}).
    
    \bibitem{Obrien2007}
    \bibinfo{author}{O'Brien, J.~L.}
    \newblock \bibinfo{title}{Optical {Quantum} {Computing}}.
    \newblock \emph{\bibinfo{journal}{Science}} \textbf{\bibinfo{volume}{318}},
    \bibinfo{pages}{1567--1570} (\bibinfo{year}{2007}).
    
    \bibitem{Dowling2008}
    \bibinfo{author}{Dowling, J.~P.}
    \newblock \bibinfo{title}{Quantum optical metrology – the lowdown on
        high-{N00N} states}.
    \newblock \emph{\bibinfo{journal}{Contemporary Physics}}
    \textbf{\bibinfo{volume}{49}}, \bibinfo{pages}{125--143}
    (\bibinfo{year}{2008}).
    
    \bibitem{Aharonovic2016}
    \bibinfo{author}{Aharonovich, I.}, \bibinfo{author}{Englund, D.} \&
    \bibinfo{author}{Toth, M.}
    \newblock \bibinfo{title}{Solid-state single-photon emitters}.
    \newblock \emph{\bibinfo{journal}{Nature Photonics}}
    \textbf{\bibinfo{volume}{10}}, \bibinfo{pages}{631--641}
    (\bibinfo{year}{2016}).
    
    \bibitem{Senellart2017}
    \bibinfo{author}{Senellart, P.}, \bibinfo{author}{Solomon, G.} \&
    \bibinfo{author}{White, A.}
    \newblock \bibinfo{title}{High-performance semiconductor quantum-dot
        single-photon sources}.
    \newblock \emph{\bibinfo{journal}{Nature Nanotechnology}}
    \textbf{\bibinfo{volume}{12}}, \bibinfo{pages}{1026--1039}
    (\bibinfo{year}{2017}).
    
    \bibitem{Arakawa2020}
    \bibinfo{author}{Arakawa, Y.} \& \bibinfo{author}{Holmes, M.~J.}
    \newblock \bibinfo{title}{Progress in quantum-dot single photon sources for
        quantum information technologies: {A} broad spectrum overview}.
    \newblock \emph{\bibinfo{journal}{Applied Physics Reviews}}
    \textbf{\bibinfo{volume}{7}}, \bibinfo{pages}{021309} (\bibinfo{year}{2020}).
    
    \bibitem{Somaschi2016}
    \bibinfo{author}{Somaschi, N.} \emph{et~al.}
    \newblock \bibinfo{title}{Near-optimal single-photon sources in the solid
        state}.
    \newblock \emph{\bibinfo{journal}{Nature Photonics}}
    \textbf{\bibinfo{volume}{10}}, \bibinfo{pages}{340--345}
    (\bibinfo{year}{2016}).
    
    \bibitem{Wang2019}
    \bibinfo{author}{Wang, H.} \emph{et~al.}
    \newblock \bibinfo{title}{Towards optimal single-photon sources from polarized
        microcavities}.
    \newblock \emph{\bibinfo{journal}{Nature Photonics}}
    \textbf{\bibinfo{volume}{13}}, \bibinfo{pages}{770--775}
    (\bibinfo{year}{2019}).
    
    \bibitem{Tomm2021}
    \bibinfo{author}{Tomm, N.} \emph{et~al.}
    \newblock \bibinfo{title}{A bright and fast source of coherent single photons}.
    \newblock \emph{\bibinfo{journal}{Nature Nanotechnology}}
    \textbf{\bibinfo{volume}{16}}, \bibinfo{pages}{399--403}
    (\bibinfo{year}{2021}).
    
    \bibitem{Wang2018}
    \bibinfo{author}{Wang, J.} \emph{et~al.}
    \newblock \bibinfo{title}{Bright room temperature single photon source at
        telecom range in cubic silicon carbide}.
    \newblock \emph{\bibinfo{journal}{Nature Communications}}
    \textbf{\bibinfo{volume}{9}}, \bibinfo{pages}{4106} (\bibinfo{year}{2018}).
    
    \bibitem{Zhou2018}
    \bibinfo{author}{Zhou, Y.} \emph{et~al.}
    \newblock \bibinfo{title}{Room temperature solid-state quantum emitters in the
        telecom range}.
    \newblock \emph{\bibinfo{journal}{Science Advances}}
    \textbf{\bibinfo{volume}{4}}, \bibinfo{pages}{eaar3580}
    (\bibinfo{year}{2018}).
    
    \bibitem{He2018}
    \bibinfo{author}{He, X.} \emph{et~al.}
    \newblock \bibinfo{title}{Carbon nanotubes as emerging quantum-light sources}.
    \newblock \emph{\bibinfo{journal}{Nature Materials}}
    \textbf{\bibinfo{volume}{17}}, \bibinfo{pages}{663--670}
    (\bibinfo{year}{2018}).
    
    \bibitem{He2017}
    \bibinfo{author}{He, X.} \emph{et~al.}
    \newblock \bibinfo{title}{Tunable room-temperature single-photon emission at
        telecom wavelengths from sp3 defects in carbon nanotubes}.
    \newblock \emph{\bibinfo{journal}{Nature Photonics}}
    \textbf{\bibinfo{volume}{11}}, \bibinfo{pages}{577--582}
    (\bibinfo{year}{2017}).
    
    \bibitem{Settele2021}
    \bibinfo{author}{Settele, S.} \emph{et~al.}
    \newblock \bibinfo{title}{Synthetic control over the binding configuration of
        luminescent sp3-defects in single-walled carbon nanotubes}.
    \newblock \emph{\bibinfo{journal}{Nature Communications}}
    \textbf{\bibinfo{volume}{12}}, \bibinfo{pages}{2119} (\bibinfo{year}{2021}).
    
    \bibitem{Kwon2016}
    \bibinfo{author}{Kwon, H.} \emph{et~al.}
    \newblock \bibinfo{title}{Molecularly {Tunable} {Fluorescent} {Quantum}
        {Defects}}.
    \newblock \emph{\bibinfo{journal}{Journal of the American Chemical Society}}
    \textbf{\bibinfo{volume}{138}}, \bibinfo{pages}{6878--6885}
    (\bibinfo{year}{2016}).
    
    \bibitem{Li2022}
    \bibinfo{author}{Li, M.-K.} \emph{et~al.}
    \newblock \bibinfo{title}{Electroluminescence from {Single}-{Walled} {Carbon}
        {Nanotubes} with {Quantum} {Defects}}.
    \newblock \emph{\bibinfo{journal}{ACS Nano}} \textbf{\bibinfo{volume}{16}},
    \bibinfo{pages}{11742--11754} (\bibinfo{year}{2022}).
    
    \bibitem{Pyatkov2016}
    \bibinfo{author}{Pyatkov, F.} \emph{et~al.}
    \newblock \bibinfo{title}{Cavity-enhanced light emission from electrically
        driven carbon nanotubes}.
    \newblock \emph{\bibinfo{journal}{Nature Photonics}}
    \textbf{\bibinfo{volume}{10}}, \bibinfo{pages}{420--427}
    (\bibinfo{year}{2016}).
    
    \bibitem{Jeantet2016}
    \bibinfo{author}{Jeantet, A.} \emph{et~al.}
    \newblock \bibinfo{title}{Widely {Tunable} {Single}-{Photon} {Source} from a
        {Carbon} {Nanotube} in the {Purcell} {Regime}}.
    \newblock \emph{\bibinfo{journal}{Physical Review Letters}}
    \textbf{\bibinfo{volume}{116}}, \bibinfo{pages}{247402}
    (\bibinfo{year}{2016}).
    
    \bibitem{Huemmer2016}
    \bibinfo{author}{Hümmer, T.} \emph{et~al.}
    \newblock \bibinfo{title}{Cavity-enhanced {Raman} microscopy of individual
        carbon nanotubes}.
    \newblock \emph{\bibinfo{journal}{Nature Communications}}
    \textbf{\bibinfo{volume}{7}}, \bibinfo{pages}{12155} (\bibinfo{year}{2016}).
    
    \bibitem{Ishii2018}
    \bibinfo{author}{Ishii, A.} \emph{et~al.}
    \newblock \bibinfo{title}{Enhanced {Single}-{Photon} {Emission} from
        {Carbon}-{Nanotube} {Dopant} {States} {Coupled} to {Silicon}
        {Microcavities}}.
    \newblock \emph{\bibinfo{journal}{Nano Letters}} \textbf{\bibinfo{volume}{18}},
    \bibinfo{pages}{3873--3878} (\bibinfo{year}{2018}).
    
    \bibitem{Luo2019}
    \bibinfo{author}{Luo, Y.} \emph{et~al.}
    \newblock \bibinfo{title}{Carbon {Nanotube} {Color} {Centers} in {Plasmonic}
        {Nanocavities}: {A} {Path} to {Photon} {Indistinguishability} at {Telecom}
        {Bands}}.
    \newblock \emph{\bibinfo{journal}{Nano Letters}} \textbf{\bibinfo{volume}{19}},
    \bibinfo{pages}{9037--9044} (\bibinfo{year}{2019}).
    
    \bibitem{Bylander2003}
    \bibinfo{author}{Bylander, J.}, \bibinfo{author}{Robert-Philip, I.} \&
    \bibinfo{author}{Abram, I.}
    \newblock \bibinfo{title}{Interference and correlation of two independent
        photons}.
    \newblock \emph{\bibinfo{journal}{The European Physical Journal D - Atomic,
            Molecular, Optical and Plasma Physics}} \textbf{\bibinfo{volume}{22}},
    \bibinfo{pages}{295--301} (\bibinfo{year}{2003}).
    
    \bibitem{Sun2009}
    \bibinfo{author}{Sun, F.~W.} \& \bibinfo{author}{Wong, C.~W.}
    \newblock \bibinfo{title}{Indistinguishability of independent single photons}.
    \newblock \emph{\bibinfo{journal}{Physical Review A}}
    \textbf{\bibinfo{volume}{79}}, \bibinfo{pages}{013824}
    (\bibinfo{year}{2009}).
    
    \bibitem{Liu2018}
    \bibinfo{author}{Liu, F.} \emph{et~al.}
    \newblock \bibinfo{title}{High purcell factor generation of indistinguishable
        on-chip single photons}.
    \newblock \emph{\bibinfo{journal}{Nature Nanotechnology}}
    \textbf{\bibinfo{volume}{13}}, \bibinfo{pages}{835--840}
    (\bibinfo{year}{2018}).
    
    \bibitem{Dusanowski2020}
    \bibinfo{author}{Dusanowski, L.} \emph{et~al.}
    \newblock \bibinfo{title}{Purcell-enhanced and indistinguishable single-photon
        generation from quantum dots coupled to on-chip integrated ring resonators}.
    \newblock \emph{\bibinfo{journal}{Nano Letters}} \textbf{\bibinfo{volume}{20}},
    \bibinfo{pages}{6357--6363} (\bibinfo{year}{2020}).
    
    \bibitem{Lakshminarayan2021}
    \bibinfo{author}{Sharma, L.} \& \bibinfo{author}{Tripathi, L.~N.}
    \newblock \bibinfo{title}{Deterministic coupling of quantum emitter to surface
        plasmon polaritons, purcell enhanced generation of indistinguishable single
        photons and quantum information processing}.
    \newblock \emph{\bibinfo{journal}{Optics Communications}}
    \textbf{\bibinfo{volume}{496}}, \bibinfo{pages}{127139}
    (\bibinfo{year}{2021}).
    
    \bibitem{Ourari2023}
    \bibinfo{author}{Ourari, S.} \emph{et~al.}
    \newblock \bibinfo{title}{Indistinguishable telecom band photons from a single
        {Er} ion in the solid state}.
    \newblock \emph{\bibinfo{journal}{Nature}} \textbf{\bibinfo{volume}{620}},
    \bibinfo{pages}{977--981} (\bibinfo{year}{2023}).
    
    \bibitem{Auffeves2010}
    \bibinfo{author}{Auffèves, A.} \emph{et~al.}
    \newblock \bibinfo{title}{Controlling the dynamics of a coupled atom-cavity
        system by pure dephasing}.
    \newblock \emph{\bibinfo{journal}{Physical Review B}}
    \textbf{\bibinfo{volume}{81}}, \bibinfo{pages}{245419}
    (\bibinfo{year}{2010}).
    
    \bibitem{Grange2015}
    \bibinfo{author}{Grange, T.} \emph{et~al.}
    \newblock \bibinfo{title}{Cavity-{Funneled} {Generation} of {Indistinguishable}
        {Single} {Photons} from {Strongly} {Dissipative} {Quantum} {Emitters}}.
    \newblock \emph{\bibinfo{journal}{Physical Review Letters}}
    \textbf{\bibinfo{volume}{114}}, \bibinfo{pages}{193601}
    (\bibinfo{year}{2015}).
    
    \bibitem{piao_brightening_2013}
    \bibinfo{author}{Piao, Y.} \emph{et~al.}
    \newblock \bibinfo{title}{Brightening of carbon nanotube photoluminescence
        through the incorporation of sp3 defects}.
    \newblock \emph{\bibinfo{journal}{Nature Chemistry}}
    \textbf{\bibinfo{volume}{5}}, \bibinfo{pages}{840--845}
    (\bibinfo{year}{2013}).
    
    \bibitem{Wang2022}
    \bibinfo{author}{Wang, P.} \emph{et~al.}
    \newblock \bibinfo{title}{Quantum defects: What pairs with the aryl group when
        bonding to the sp2 carbon lattice of single-wall carbon nanotubes?}
    \newblock \emph{\bibinfo{journal}{Journal of the American Chemical Society}}
    \textbf{\bibinfo{volume}{144}}, \bibinfo{pages}{13234--13241}
    (\bibinfo{year}{2022}).
    
    \bibitem{Weisman2003}
    \bibinfo{author}{Weisman, R.~B.} \& \bibinfo{author}{Bachilo, S.~M.}
    \newblock \bibinfo{title}{Dependence of optical transition energies on
        structure for single-walled carbon nanotubes in aqueous suspension: An
        empirical kataura plot}.
    \newblock \emph{\bibinfo{journal}{Nano Letters}} \textbf{\bibinfo{volume}{3}},
    \bibinfo{pages}{1235--1238} (\bibinfo{year}{2003}).
    
    \bibitem{ITU2016}
    \bibinfo{author}{ITU-T}.
    \newblock \bibinfo{title}{Optical system design and engineering
        considerations}.
    \newblock \bibinfo{type}{Tech. Rep.} \bibinfo{number}{ITU-T G Suppl. 39},
    \bibinfo{institution}{International Telecommunication Union}
    (\bibinfo{year}{2016}).
    
    \bibitem{Hunger2010}
    \bibinfo{author}{Hunger, D.} \emph{et~al.}
    \newblock \bibinfo{title}{A fiber fabry–perot cavity with high finesse}.
    \newblock \emph{\bibinfo{journal}{New Journal of Physics}}
    \textbf{\bibinfo{volume}{12}}, \bibinfo{pages}{065038}
    (\bibinfo{year}{2010}).
    
    \bibitem{Kaupp2013}
    \bibinfo{author}{Kaupp, H.} \emph{et~al.}
    \newblock \bibinfo{title}{Scaling laws of the cavity enhancement for
        nitrogen-vacancy centers in diamond}.
    \newblock \emph{\bibinfo{journal}{Phys. Rev. A}} \textbf{\bibinfo{volume}{88}},
    \bibinfo{pages}{053812} (\bibinfo{year}{2013}).
    
    \bibitem{Hoegele_photon_2008}
    \bibinfo{author}{Högele, A.}, \bibinfo{author}{Galland, C.},
    \bibinfo{author}{Winger, M.} \& \bibinfo{author}{Imamoğlu, A.}
    \newblock \bibinfo{title}{Photon {Antibunching} in the {Photoluminescence}
        {Spectra} of a {Single} {Carbon} {Nanotube}}.
    \newblock \emph{\bibinfo{journal}{Physical Review Letters}}
    \textbf{\bibinfo{volume}{100}}, \bibinfo{pages}{217401}
    (\bibinfo{year}{2008}).
    
    \bibitem{Hartschuh2003}
    \bibinfo{author}{Hartschuh, A.}, \bibinfo{author}{Pedrosa, H.~N.},
    \bibinfo{author}{Novotny, L.} \& \bibinfo{author}{Krauss, T.~D.}
    \newblock \bibinfo{title}{Simultaneous fluorescence and raman scattering from
        single carbon nanotubes}.
    \newblock \emph{\bibinfo{journal}{Science}} \textbf{\bibinfo{volume}{301}},
    \bibinfo{pages}{1354--1356} (\bibinfo{year}{2003}).
    
    \bibitem{Auffeves2009}
    \bibinfo{author}{Auffèves, A.}, \bibinfo{author}{Gérard, J.-M.} \&
    \bibinfo{author}{Poizat, J.-P.}
    \newblock \bibinfo{title}{Pure emitter dephasing: {A} resource for advanced
        solid-state single-photon sources}.
    \newblock \emph{\bibinfo{journal}{Physical Review A}}
    \textbf{\bibinfo{volume}{79}}, \bibinfo{pages}{053838}
    (\bibinfo{year}{2009}).
    
    \bibitem{Santori2002}
    \bibinfo{author}{Santori, C.}, \bibinfo{author}{Fattal, D.},
    \bibinfo{author}{Vučković, J.}, \bibinfo{author}{Solomon, G.~S.} \&
    \bibinfo{author}{Yamamoto, Y.}
    \newblock \bibinfo{title}{Indistinguishable photons from a single-photon
        device}.
    \newblock \emph{\bibinfo{journal}{Nature}} \textbf{\bibinfo{volume}{419}},
    \bibinfo{pages}{594--597} (\bibinfo{year}{2002}).
    
    \bibitem{Hartmann2016}
    \bibinfo{author}{Hartmann, N.~F.} \emph{et~al.}
    \newblock \bibinfo{title}{Photoluminescence {Dynamics} of {Aryl} sp3 {Defect}
        {States} in {Single}-{Walled} {Carbon} {Nanotubes}}.
    \newblock \emph{\bibinfo{journal}{ACS Nano}} \textbf{\bibinfo{volume}{10}},
    \bibinfo{pages}{8355--8365} (\bibinfo{year}{2016}).
    
    \bibitem{Kaupp2016}
    \bibinfo{author}{Kaupp, H.} \emph{et~al.}
    \newblock \bibinfo{title}{Purcell-enhanced single-photon emission from
        nitrogen-vacancy centers coupled to a tunable microcavity}.
    \newblock \emph{\bibinfo{journal}{Phys. Rev. Appl.}}
    \textbf{\bibinfo{volume}{6}}, \bibinfo{pages}{054010} (\bibinfo{year}{2016}).
    
    \bibitem{Knill2001}
    \bibinfo{author}{Knill, E.}, \bibinfo{author}{Laflamme, R.} \&
    \bibinfo{author}{Milburn, G.~J.}
    \newblock \bibinfo{title}{A scheme for efficient quantum computation with
        linear optics}.
    \newblock \emph{\bibinfo{journal}{Nature}} \textbf{\bibinfo{volume}{409}},
    \bibinfo{pages}{46--52} (\bibinfo{year}{2001}).
    
    \bibitem{Azuma2015}
    \bibinfo{author}{Azuma, K.}, \bibinfo{author}{Tamaki, K.} \&
    \bibinfo{author}{Lo, H.-K.}
    \newblock \bibinfo{title}{All-photonic quantum repeaters}.
    \newblock \emph{\bibinfo{journal}{Nature Communications}}
    \textbf{\bibinfo{volume}{6}}, \bibinfo{pages}{6787} (\bibinfo{year}{2015}).
    
    \bibitem{Zheng2015}
    \bibinfo{author}{Fagan, J.~A.} \emph{et~al.}
    \newblock \bibinfo{title}{Isolation of >1 nm diameter single-wall carbon
        nanotube species using aqueous two-phase extraction}.
    \newblock \emph{\bibinfo{journal}{ACS Nano}} \textbf{\bibinfo{volume}{9}},
    \bibinfo{pages}{5377--5390} (\bibinfo{year}{2015}).
    
    \bibitem{Hunger2012}
    \bibinfo{author}{Hunger, D.}, \bibinfo{author}{Deutsch, C.},
    \bibinfo{author}{Barbour, R.~J.}, \bibinfo{author}{Warburton, R.~J.} \&
    \bibinfo{author}{Reichel, J.}
    \newblock \bibinfo{title}{Laser micro-fabrication of concave, low-roughness
        features in silica}.
    \newblock \emph{\bibinfo{journal}{AIP Advances}} \textbf{\bibinfo{volume}{2}},
    \bibinfo{pages}{012119} (\bibinfo{year}{2012}).
    
    \bibitem{Fischer2016}
    \bibinfo{author}{Fischer, K.~A.}, \bibinfo{author}{Müller, K.},
    \bibinfo{author}{Lagoudakis, K.~G.} \& \bibinfo{author}{Vučković, J.}
    \newblock \bibinfo{title}{Dynamical modeling of pulsed two-photon
        interference}.
    \newblock \emph{\bibinfo{journal}{New Journal of Physics}}
    \textbf{\bibinfo{volume}{18}}, \bibinfo{pages}{113053}
    (\bibinfo{year}{2016}).
    
\end{thebibliography}
\end{document}


\title{Supplementary Information:\\ Cavity-enhanced photon indistinguishability \\ at room temperature and telecom wavelengths}

\author{Lukas Husel}
\altaffiliation{These authors contributed equally to this work}
\def\LMU{Fakult\"at f\"ur Physik, Munich Quantum Center, and Center for NanoScience (CeNS), Ludwig-Maximilians-Universit\"at M\"unchen, Geschwister-Scholl-Platz 1, 80539 M\"unchen, Germany}
\affiliation{\LMU}

\author{Julian Trapp}
\altaffiliation{These authors contributed equally to this work}
\affiliation{\LMU}

\author{Johannes Scherzer}
\affiliation{\LMU}

\author{Xiaojian Wu}
\def\MRY{Department of Chemistry and Biochemistry, University of Maryland, Maryland, USA}
\affiliation{\MRY}

\author{Peng Wang}
\affiliation{\MRY}

\author{Jacob Fortner}
\affiliation{\MRY}

\author{Manuel Nutz}
\def\QLIBRI{Qlibri GmbH, Maistr. 67, 80337 M\"unchen, Germany}
\affiliation{\QLIBRI}

\author{Thomas Hümmer}
\affiliation{\QLIBRI}

\author{Borislav Polovnikov}
\affiliation{\LMU}

\author{Michael F\"org}
\def\QLIBRI{Qlibri GmbH, Maistr. 67, 80337 M\"unchen, Germany}
\affiliation{\QLIBRI}

\author{David Hunger}
\affiliation{Physikalisches Institut, Karlsruhe Institute of Technology, Karlsruhe, Germany}  
\affiliation{Institute for Quantum Materials and Technologies (IQMT), Karlsruhe Institute of Technology (KIT),  Herrmann-von-Helmholtz Platz 1, 76344 Eggenstein-Leopoldshafen, Germany}

\author{YuHuang Wang}
\affiliation{\MRY}

\author{Alexander H\"ogele}
\affiliation{\LMU}
\affiliation{Munich Center for Quantum Science and Technology (MCQST), Schellingstr. 4, 80799 M\"unchen, Germany}

\maketitle

\vspace{-4pt}
\section*{Supplementary Note 1 - Regime of incoherent good NTD-cavity coupling}

\label{sec:incReg}
In the regime of incoherent good cavity coupling, ${2g \ll \gamma + \gamma^* + \kappa}$ and ${\kappa < \gamma + \gamma^*}$ holds~\cite{Grange2015} , where $g$ is the light-matter coupling strength, $\gamma$ is the population decay rate, $\kappa$ is the cavity linewidth and $\gamma^*$ is the pure dephasing rate. $\gamma = \gamma_r + \gamma_{nr}$ is the sum of radiative and nonradiative decay rates $\gamma_r$ and $\gamma_{nr}$, respectively. We determined ${\kappa = 35.4 \pm 0.1~\mu}$eV for the lowest accessible longitudinal mode order, and ${\gamma^* = 8 \pm 2}$~meV from the cavity length sweep in Fig.~2b of the main text. As pointed out ibidem, our two-photon interference and PL lifetime measurements indicate a biexponential NTD population decay with fast and slow timescales $\tau_{\mathrm{fast}}=2$~ps and $\tau_{\mathrm{slow}}=91$~ps, respectively, corresponding to ${\gamma_{\mathrm{fast}} = 330~\mu}$eV and ${\gamma_{\mathrm{slow}} = 7.3~\mu}$eV. Obviously, ${\kappa < \gamma + \gamma^*}$ holds in our system for both $\gamma=\gamma_{\mathrm{fast}}$ and $\gamma=\gamma_{\mathrm{slow}}$.

We estimate the light-matter coupling strength using 
\begin{equation}
    g = \sqrt{3\lambda^2c/(8\pi n^3 V_{\mathrm{c}}\tau_{\mathrm{rad}})},
    \label{Eq:g}    
\end{equation}
with the speed of light $c$, the wavelength $\lambda$, the cavity mode volume $V_{\mathrm{c}}$, the refractive index $n$ and the radiative lifetime $\tau_{\mathrm{rad}}$ \cite{Hunger2010}. For our cavity, we calculated $V_{\mathrm{c}} = 8.2~\mu \mathrm{m}^3$ for the lowest accessible mode order. We use $n=1$ in our estimate of the light-matter coupling strength, neglecting that the NTDs are placed on a polystyrene spacer with refractive index $n = 1.57$. This results in an upper bound for $g$. For the type of NTDs investigated in this work, $\tau_{\mathrm{rad}}$ was found to range between 1 and 15~ns \cite{He2017, Hartmann2016}. Varying $\tau_{rad}$ within $1-15$~ns, we expect $g$ to range between 17 and 64~$\mu$eV. With the results above, we conclude that ${2g \ll \gamma + \gamma^* + \kappa}$ holds in our system, consistent with the regime of incoherent good cavity coupling. 

\section*{Supplementary Note 2 - Model for NTD-cavity coupling dynamics} 

We first consider a two-level emitter coupled to an optical cavity. This setting was studied in Ref.~\cite{Grange2015}, where the time-dependent density operator $\hat{\rho}(t)$ of the coupled emitter-cavity system was obtained from a Lindblad master equation in Markovian approximation. In the regime of incoherent cavity coupling, the influence of the density operator coherences on the system dynamics was shown to be negligible. The coupled system is therefore fully described by the populations of cavity and emitter, which exchange photons at a rate $R$ given by~\cite{Auffeves2009}:
\begin{equation}
    R = 4g^2/(\kappa+\gamma+\gamma^*).
    \label{Eq:R}
\end{equation}

As explained in the main text, our two-photon interference and PL lifetime measurements indicate a biexponential NTD population decay, attributed to the presence of an additional dark excitonic reservoir. In order to model the coupling of such an NTD to a cavity, we first focus on the value of $R$ in our system. Given incoherent cavity coupling, the denominator in Eq.~\ref{Eq:R} is dominated by the pure dephasing rate $\gamma^*$, such that $R \approx 4g^2/\gamma^*$. From the measured value ${\gamma^* = 8 \pm 2}$~meV and the estimated range for $g$ ($17-64~\mu$eV), we expect $R$ to range between 0.14 and 2.04~$\mu$eV.

We now extend the model of Ref.~\cite{Grange2015} to an NTD exhibiting dark and bright exciton states. The dark state has no effect on the limit of incoherent coupling due to its vanishingly small coupling to the cavity. The bright state, on the other hand, will exchange photons with the cavity at rate $R$, analogous to a radiative two-level system. Based on these considerations, we describe our experiment with the set of partial differential equations: 
\begin{align}
    \label{Eq:cavitypop1}
    &\frac{\partial \rho_{\mathrm{c}}}{\partial t} = -(\kappa + R) \rho_{\mathrm{c}} + R\rho_{\mathrm{b}} \\
    &\frac{\partial \rho_{\mathrm{d}}}{\partial t} = -\gamma_d \rho_{\mathrm{d}} + \gamma_{\mathrm{bd}} \rho_{\mathrm{b}} - \gamma_{\mathrm{db}} \rho_{\mathrm{d}} \\
    &\frac{\partial \rho_{\mathrm{b}}}{\partial t} = -(\gamma_b + R)\rho_{\mathrm{b}} + R\rho_{\mathrm{c}} - \gamma_{\mathrm{bd}} \rho_{\mathrm{b}} + \gamma_{\mathrm{db}} \rho_{\mathrm{d}}
    \label{Eq:cavitypop2}   
\end{align}
with the populations of the cavity $\rho_{\mathrm{c}}$, dark state $\rho_{\mathrm{d}}$ and bright state $\rho_{\mathrm{b}}$. Here, $\gamma_{\mathrm{bd/db}}$ is the population exchange rate between $\rho_{\mathrm{b}}$, and $\rho_{\mathrm{d}}$ and $\gamma_{\mathrm{b/d}}$ are the sum of radiative and nonradiative decay rates of $\rho_{\mathrm{b}}$ and $\rho_{\mathrm{d}}$. 

The parameters $\gamma_{\mathrm{b/d}}$, $\gamma_{\mathrm{bd/db}}$ and $\rho_b(0)$ are free parameters in our model. We choose them such that our model prediction fulfils two criteria. First, the predicted time-dependent PL intensity, which is the cavity population $\rho_{\mathrm{c}}$ convoluted with the instrument response function, should agree with the measured time-dependent PL in Fig.~4d of the main text. Second, the biexponential bright state decay, obtained for setting $R=0$ in Eqns.~\ref{Eq:cavitypop1}~--~\ref{Eq:cavitypop2}, should have a short population lifetime $\tau_{\mathrm{short}}=2$~ps. This is indicated by the two-photon interference measurement in Fig.~4b of the main text, as explained ibidem and in Supplementary Note 5. The result of this parameter adaptation is the solid line in Fig.~4d and agrees well with the measured data.

\section*{Supplementary Note 3 - Regime of low Purcell enhancement}

\vspace{-8pt}
\subsection{Purcell factor and PL lifetime}

The PL lifetime $\tau_{\mathrm{pl}}$ of emitters coupled to the cavity is given by ${\tau_{\mathrm{pl}} = \tau_{\mathrm{fs}}/(1+F_{\mathrm{p}}^*)}$, with the free-space population lifetime $\tau_{\mathrm{fs}} = 1/\gamma$ and the effective Purcell factor $F_{\mathrm{p}}^* = R/\gamma$ \cite{Auffeves2010}. For the slow population decay component with $1/\gamma = 91$~ps, $F_{\mathrm{p}}^*$ ranges between 0.018 and 0.28 for the previously estimated range of $R$ ($0.14 - 2.04~\mu$eV). For the fast decay component, the expected value for $F_{\mathrm{p}}^*$ is even smaller ($4.1\cdot10^{-4} - 6.2\cdot10^{-3}$). We infer from this result that population lifetime shortening effects due to the cavity coupling are negligible in our system, corresponding to a regime of low Purcell enhancement. 

While $F_{\mathrm{p}}^*$ quantifies enhancement of the total emitter decay rate by cavity-coupling, the enhancement of the radiative decay rate $\gamma_{\mathrm{rad}} = 1/\tau_{\mathrm{rad}}$ is quantified by $F_{\mathrm{p}} = 3\lambda^3Q_{\mathrm{eff}}/(4\pi^2n^3V_c)$, with the effective Q-factor $Q_{\mathrm{eff}} = (Q_{\mathrm{cav}}^{-1}+Q_{\mathrm{em}}^{-1})^{-1}$ and the Q-factors of cavity $Q_{\mathrm{cav}}$ and emitter $Q_{\mathrm{em}}$, respectively. In the limit of incoherent coupling, $F_{\mathrm{p}}$ approximates to $F_{\mathrm{p}} \approx R/\gamma_{\mathrm{rad}} \approx 3\lambda^2c/(4\pi V_{\mathrm{c}}\gamma^*)$ \cite{Jeantet2016}. For our system, we expect $F_{\mathrm{p}} = 1.6$. We note that for $\kappa \gg \gamma + \gamma^*$ (bad cavity regime, different from our experiment), $F_{\mathrm{p}}$ is called ideal Purcell factor~\cite{Benedikter2017} and evaluates to the original expression given by Purcell~\cite{Purcell1946} $F_{\mathrm{p}} \approx 3\lambda^3Q_{\mathrm{c}}/(4\pi^2n^3V_c) = 4g^2/(\gamma_{\mathrm{rad}}\kappa)$. Finally, we also note that the effective Purcell factor can be expressed as $F_{\mathrm{p}}^* = \eta_Q F_{\mathrm{p}}$, with the quantum yield $\eta_Q = \gamma_r/\gamma$. From the expected range of radiative lifetimes given above, we infer an estimated range of $6\cdot10^{-3} - 9\cdot10^{-2}$ for $\eta_Q$, where we only considered the fast decay component since the majority of the population decays on this timescale (see Supplementary Note 4). The estimated values agree with previously measured NTD quantum yields.

\vspace{-8pt}
\subsection{Single photon efficiency}

The single photon emission efficiency $\beta_{\mathrm{c}}$ gives the probability that a photon is emitted into the spectral window of the cavity linewidth $\kappa$, given an initial excitation of the emitter. Each excitation pulse generates a photon in the cavity with probability $\beta_{\mathrm{c}}\eta_{\mathrm{em}}$, where $\eta_{\mathrm{em}}$ is the free space photon emission efficiency at the respective pump power. The rate of photons registered by the detector $I_{\mathrm{em}}$ is then given by:

\begin{equation}
    I_{\mathrm{em}} = f_{\mathrm{exc}}\eta_{\mathrm{out}}\eta_{\mathrm{sys}}\eta_{\mathrm{em}}\beta_{\mathrm{c}},
    \label{I_em}
\end{equation} \\
\noindent where $f_{\mathrm{exc}}$ is the repetition rate of the excitation source, $\eta_{\mathrm{out}}$ is the probability for a photon to exit the cavity through the flat mirror, $\eta_{\mathrm{sys}}$ is the combined transmission and detection efficiency of the setup.

The measurement of $\eta_{\mathrm{em}}$ would require excitation near or above the saturation threshold. This in turn requires high excitation powers, which can lead to NTD degradation and limit single photon purity and indistinguishability. Since NTD~1, 2 and 3 were used to benchmark single photon purity and indistinguishability, we refrained from measurements at such excitation powers. However, using the maximum measured value $I_{\mathrm{em}}= 1840 \pm 30$ counts/s for NTD1 and the upper bound $\eta_{\mathrm{em}}=1$, combined with the measurement of $\eta_{\mathrm{sys}}$ and the value for $\eta_{\mathrm{out}}$ obtained from transfer matrix simulations of the mirror coating, we obtain a lower bound of $\mathrm{min}(\beta_{\mathrm{c}}) = (3.9 \pm 0.1) \cdot 10^{-3}$ for the single photon emission efficiency. The theoretically expected value is calculated from the time-dependent cavity population $\rho_{\mathrm{c}}$ as~\cite{Grange2015}:
\begin{equation}
    \beta_{\mathrm{c}} = \kappa \int \rho_{\mathrm{c}}(t) dt.
    \label{beta_theo}
\end{equation}
Solving Eqns.~\ref{Eq:cavitypop1}~--~\ref{Eq:cavitypop2} for $\rho_{c}$, we obtain an expected value $\beta_{\mathrm{c}} = 6.6 \cdot 10^{-3}$ for the single photon emission efficiency.

In the main text, we compared the above values to the expected upper bound for the emission efficiency for spectrally filtered free-space emission ${\beta_{\mathrm{fs}}}$, similar to Ref.~\cite{Grange2015}. The actual value for this quantity is likely smaller due to the non-unity quantum yield in our system, which we expect to reduce $\beta_{\mathrm{fs}}$ to $\eta_Q \kappa/(\pi\gamma^*)$. This expression is easily understood by noting that $\beta_{\mathrm{fs}}$ gives the probability that a photon is emitted into the spectral window of a filter with bandwidth $\kappa$, given an initial excitation of the emitter. Using the estimated range for $\eta_Q$ given above, we expect that the single-photon emission efficiency in our system outperforms that expected for filtered free space emission by at least a factor of 44. This drastic increase in emission spectral density is a direct consequence of cavity-coupling in the good cavity regime and was found to be quantified by the ideal Purcell factor $F_{\mathrm{p, ideal}} = 3\lambda^3Q_{\mathrm{c}}/(4\pi^2n^3V_c)$ as defined in Supplementary~Note~3.A~\cite{Kaupp2013}, for which we expect $F_{\mathrm{p, ideal}} = 91$ for the parameters given in Supplementary~Note~1.

We also use the measured single photon efficiency to confirm the result for the light-matter coupling strength $g$ from Supplementary~Note~1. If defined as in Eq.~\ref{I_em}, the efficiency is given by $\beta_{\mathrm{c}} = F_{\mathrm{p}}/(F_{\mathrm{p}} + \gamma/\gamma_{\mathrm{rad}})$ \cite{Benedikter2017}. We note that this expression is valid for all regimes of cavity-coupling. From the measured values of slow population decay lifetime and $\beta_{\mathrm{c}}$, we find an experimental upper bound $\mathrm{max}(\tau_{\mathrm{rad}}) = 35.0 \pm 0.9$~ns for the radiative lifetime, corresponding to $\mathrm{min}(g) = 13.5 \pm 0.3 \ \mu$eV. A more realistic value is obtained by also considering the fast population decay component. Since $\tau_{\mathrm{fast}} \ll \tau_{\mathrm{slow}}$, both decay processes contribute to the overall efficiency on different timescales, such that we can approximate $\beta_{\mathrm{c}} \approx \tilde{A}_{\mathrm{fast}}F_{\mathrm{p}}/(F_{\mathrm{p}} + \gamma_{\mathrm{fast}}/\gamma_{\mathrm{rad}}) + \tilde{A}_{\mathrm{slow}}F_{\mathrm{p}}/(F_{\mathrm{p}} + \gamma_{\mathrm{slow}}/\gamma_{\mathrm{rad}})$, with the fractional amplitudes $\tilde{A}_{\mathrm{fast/slow}}$ defined as in Eq.~\ref{Eq:Atilde}. We find $\tau_{\mathrm{rad}} = 12.3 \pm 0.3$~ns for the radiative lifetime, corresponding to $g = 22.8 \pm 0.6 \ \mu$eV, in agreement with the result from Supplementary~Note~1.

\vspace{-8pt}
\subsection{Cavity-enhancement of PL intensity}

From Eq.~\ref{I_em}, we derive a quantitative description of the increase in the PL intensity observed in Fig.~2c of the main text as the cavity length is tuned to the lowest accessible mode order. The single photon emission efficiency is given by ${\beta_{\mathrm{c}} = \kappa R/[\kappa R + \gamma(\kappa + R)]}$~\cite{Auffeves2009}, which in the Purcell regime ($R\ll \kappa$, as is the case for our system) simplifies to ${\beta_{\mathrm{c}} \approx R/\gamma/(1+R/\gamma) = F_{\mathrm{p}}^*/(1+F_{\mathrm{p}}^*)}$. Combined with Eqns.~\ref{Eq:g} and \ref{Eq:R}, we find $I_{\mathrm{em}} \propto R \propto V_{\mathrm{c}}^{-1}$ to first order in $R$, which quantitatively describes the behaviour observed in Fig.~2c. We conclude that in our system, cavity coupling increases the single photon emission efficiency $\beta_{\mathrm{c}}$ via enhancement of light-matter coupling strength. 

\section*{Supplementary Note 4 - Two-photon interference visibility} 

\begin{figure}[t]
    \centering
    \includegraphics[scale=1]{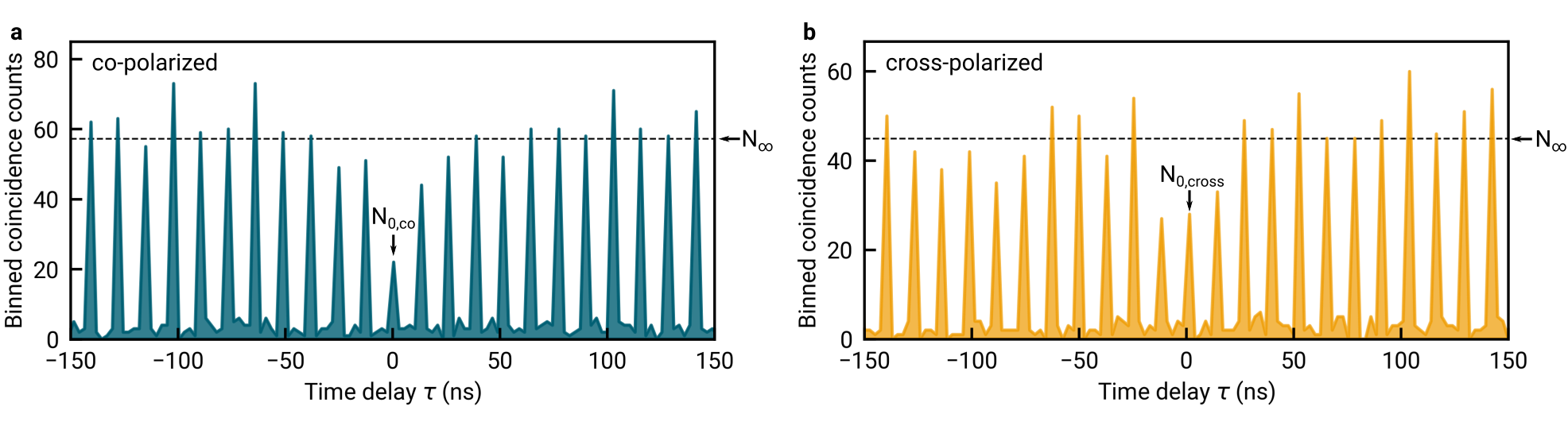}
    \caption{\textbf{HOM autocorrelation histograms for NTD3.} HOM correlations for NTD~3 for co-polarized (\textbf{a}) and cross-polarized (\textbf{b}) interferometer arms with delay of one excitation pulse. Coincidence counts were binned in 2.5~ns time windows. The HOM autocorrelation function in Fig.~4b of the main text was calculated from the displayed data (see the Methods section for details).}
    \label{figS1}
\end{figure}

We quantify the indistinguishability of photons emitted by the NTD-cavity system by the two-photon interference visibility $v$ one would obtain in an interferometer with balanced beamsplitters (BSs) and unity classical visibility. To determine $v$ from the experimental correlation histograms, we account both for imbalanced interferometer arms and non-ideal single photon purity of each NTD. 

A fiber-based interferometer as in Fig.~4a of the main text was used to perform two-photon interference experiments. The stream of photons entering the interferometer was divided at BS~1 with transmission and reflection $T_{1}$ and $R_{2}$, respectively, and recombined after a tunable delay at BS~2 with transmission and reflection $T_2$ and $R_2$. In our interferometer, the transmission of the delay arm is $\mu <$~1. The delay time equals the excitation pulse separation, and is orders of magnitude larger than the coherence time and population lifetimes in our system. We therefore treat the reduced transmission in the delay arm as an effect of imbalanced transmission and reflection of BS~1, and use the effective values $\tilde{T_1} = T_{\mathrm{1}}/(T_{\mathrm{1}}+\mu R_{\mathrm{1}})$ and $\tilde{R_1} = \mu R_{\mathrm{1}}/(T_{\mathrm{1}}+\mu R_{\mathrm{1}})$ for its transmission and reflection.

Using effective transmission and reflection values, we derive expressions for the integrated peak counts $N$ in the experimental histograms (as shown in Supplementary~Fig.~1 for NTD~3) obtained by integrating correlation events in a $2.5$~ns time window. We extend the calculation of Ref.~\cite{Kiraz2004} to obtain $N$ from the intensity autocorrelation between the output ports of BS~2. For large $|\tau|$, we find 
\begin{equation}
    N_{\mathrm{\infty}} = H\left[\tilde{R_1} \tilde{T_1} (R_2^2 + T_2^2) + R_2 T_2 (\tilde{R_1}^2 + \tilde{T_1}^2)\right],
    \label{Ninfty} 
\end{equation}
with an integration constant $H$. 

For the height of the peak at $\tau=0$ for co-polarized interferometer arms (Supplementary~Fig.~1a), we find
\begin{equation}
    N_{0, co} = H \left[\vphantom{g_{\mathrm{HBT}}^{(2)}(0)}\tilde{R_1} \tilde{T_1}\left[1 - 2 R_2 T_2 - 2 R_2 T_2 (\epsilon_P)^2 v\right]   
 + g_{\mathrm{HBT}}^{(2)}(0) R_2 T_2 (1 - 2 \tilde{R_1} \tilde{T_1})\right].
    \label{N0}
\end{equation}
In this expression, $\epsilon_P$ is the overlap between the polarization modes of the interferometer arms. The height of the central histogram peak for cross-polarization $N_{0, cross}$ (Supplementary~Fig.~1b) is obtained by setting $v=0$ in Eq.~\ref{N0}, which yields
\begin{equation}
    N_{\mathrm{0, cross}} = H \left[\vphantom{g_{\mathrm{HBT}}^{(2)}(0)}\tilde{R_1} \tilde{T_1}\left(1 - 2 R_2 T_2 \right)   
 + g_{\mathrm{HBT}}^{(2)}(0) R_2 T_2 (1 - 2 \tilde{R_1} \tilde{T_1})\right].
    \label{N0cross}
\end{equation}

To determine $v$ for NTD~3, we first extracted the raw visibility ${v_{\mathrm{raw}} = 1- g^{(2)}_{\mathrm{HOM, co}}(0)/g^{(2)}_{\mathrm{HOM, cross}}(0)}$ from the data in Fig~4b. Next, we calculated $v_{\mathrm{raw}} = 1 - N_{\mathrm{0, co}}/N_{\mathrm{0, cross}}$ from Eqns.~\ref{N0} and \ref{N0cross} and solved for $v$, yielding an expression which depends on $v_{\mathrm{raw}}$, $g^{(2)}_{\mathrm{HBT}}(0)$ and the interferometer parameters $\tilde{T_1}$, $\tilde{R_1}$, $T_2$, $R_2$ and $\epsilon_P$. Using experimental values for these quantities ($\tilde{T_1} = 0.4$, $\tilde{R_1}=0.6$, $T_2=0.49$, $R_2=0.51$ and $\epsilon_P=0.96$), we finally obtained $v = 0.51 \pm 0.21$ for NTD~3. Based on the experimental values $\tilde{T_1}$, $\tilde{R_1}$, $T_2$, $R_2$, $\epsilon_P$ and $g^{(2)}_{\mathrm{HBT}}(0)$ for NTD~3, we estimate ${g^{(2)}_{\mathrm{HOM, cross}}(0) = N_{\mathrm{0, cross}}/N_{\mathrm{\infty}}=0.53 \pm 0.04}$ for cross-polarized interferometer arms, in good agreement with the experimental value of $0.61 \pm 0.12$.

For NTD~1, we obtained the raw visibility $v_{\mathrm{raw}} = a$ from the amplitude $a$ of the best-fit to the HOM dip in Fig.~4d of the main text, and calculated $v$ from $v_{\mathrm{raw}}$, $g^{(2)}_{\mathrm{HBT}}(0)$, and the interferometer parameters as described above to obtain  $v=0.65 \pm 0.24$ stated in the main text. The observed asymmetric increase in the visibility towards large positive delays is the result of a degradation-induced decrease in the single photon purity during the measurement. For large interferometer delays, we expect \mbox{$g^{(2)}_{\mathrm{HOM}}(0) = N_{\mathrm{0, cross}}/N_{\mathrm{\infty}}=0.64 \pm 0.05$} as an estimate for the offset $c$ in the data of Fig.~4d. This value is smaller than the best-fit value $c= 0.80 \pm 0.08$, which could stem from the degradation observed during the measurement resulting in an overall increase in $g^{(2)}_{\mathrm{HOM}}$. In Supplementary~Fig.~2, we show the autocorrelation histograms for interferometer delays of -5, -3, 0, 3 and 5 ps, respectively. The reduction of correlation events at zero time delay $\tau$ provides evidence for two-photon interference for NTD~1. Each data point in Fig.~4d of the main text was obtained by taking such histograms and computing $g^{(2)}_{\mathrm{HOM}}(0)$ as described above.

\begin{figure}[t!]
    \centering
    \includegraphics[scale=1]{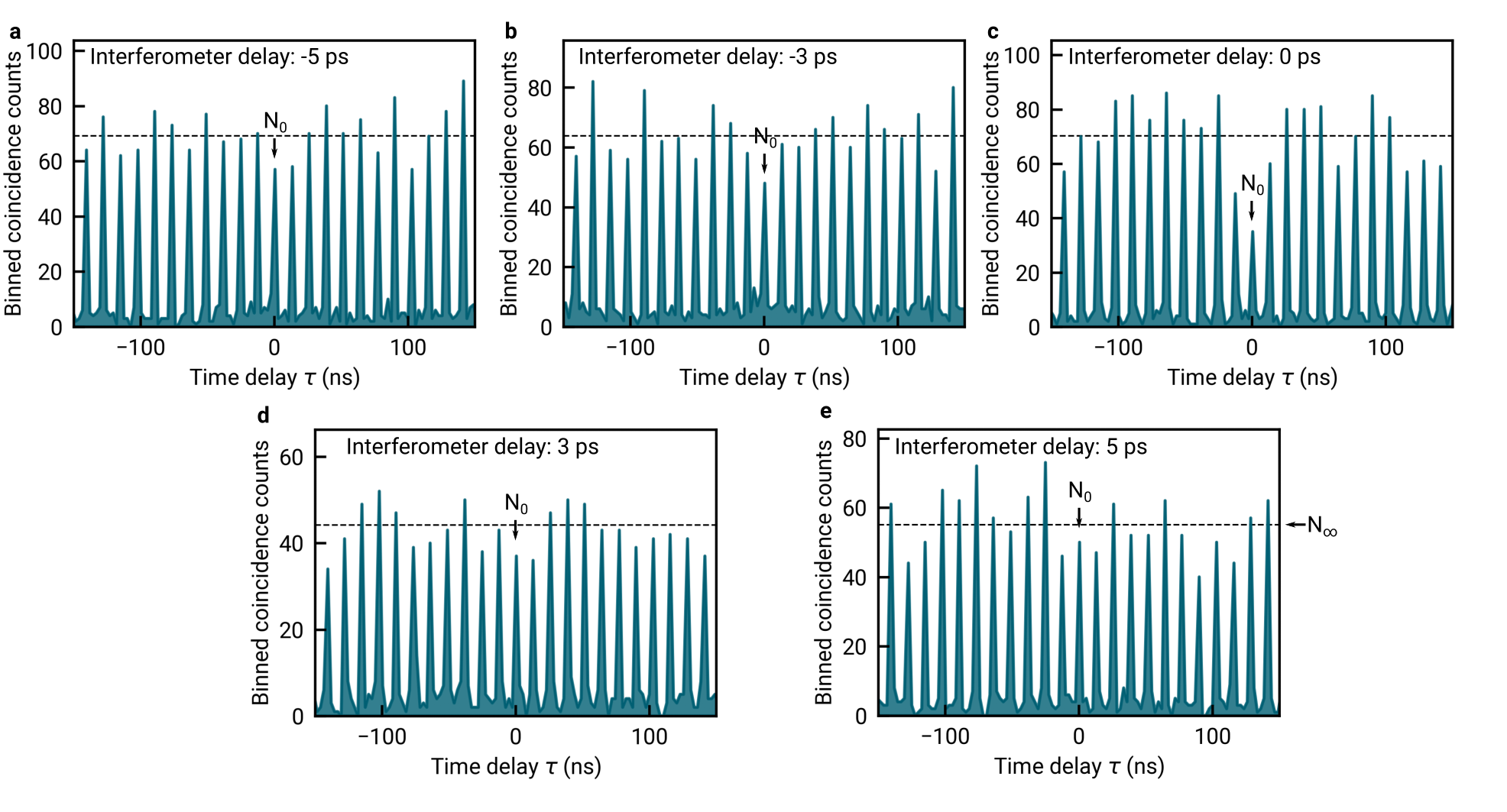}
    \caption{\textbf{HOM autocorrelation histograms for NTD~1.} \textbf{a}--\textbf{e}, HOM autocorrelation function measured on NTD1 for interferometer delays -5 ps (\textbf{a}), -3 ps (\textbf{b}), 0 ps (\textbf{c}), 3 ps (\textbf{d}) and 5 ps (\textbf{e}), as used to extract the data points of the HOM dip in Fig.~4c of the main text. The dashed line indicates the mean value of the histogram peaks at time delays $|\tau|>12.5$~ns, $N_{\mathrm{\infty}}$. An interferometer delay of 0 ps corresponds to separation by exactly one excitation pulse, resulting in maximum probability for two-photon interference and reduced correlation events at zero time delay $\tau$.}
    \label{figS2}
\end{figure}

For completeness, we also calculate the values of the HOM autocorrelation function at time delays corresponding to one excitation pulse separation, $\tau = \pm 12.5$~ns. As explained e.g. in the supplement of Ref.~\cite{Loredo2016}, $g^{(2)}_{\mathrm{HOM}}$ is smaller than one for these delays. For the respective histogram peak heights, we find
\begin{equation}
    N_{\mathrm{12.5}} = H \left[\vphantom{g_{\mathrm{HBT}}^{(2)}(0)}\tilde{R_1} \tilde{T_1} T_2^2  + R_2 T_2 (1 - 2 \tilde{R_1} \tilde{T_1}) 
+ g_{\mathrm{HBT}}^{(2)}(0) \tilde{R_1} \tilde{T_1} R_2^2\right]
    \label{N0}
\end{equation}
and
\begin{equation}
    N_{\mathrm{-12.5}} = H \left[\vphantom{g_{\mathrm{HBT}}^{(2)}(0)}\tilde{R_1} \tilde{T_1} R_2^2  + R_2 T_2 (1 - 2 \tilde{R_1} \tilde{T_1}) 
    + g_{\mathrm{HBT}}^{(2)}(0) \tilde{R_1} \tilde{T_1} T_2^2\right].
    \label{N0cross}
\end{equation}
We note that these expressions are valid for both co- and cross-polarized interferometer arms since the excitation pulse separation greatly exceeds the photon coherence time, leading to vanishing contributions of quantum interference at these time delays. From our measured values of the interferometer parameters, we calculate the expected value \mbox{$g^{(2)}_{\mathrm{HOM}}(-12.5~\mathrm{ns}) = N_{\mathrm{-12.5}}/N_{\mathrm{\infty}}=0.77 \pm 0.02$}, in agreement with the value $0.87 \pm 0.13$ measured in co-polarized configuration and close to the value $0.58 \pm 0.12$ measured in cross-polarized configuration. We also expect \mbox{$g^{(2)}_{\mathrm{HOM}}(12.5~\mathrm{ns}) = N_{\mathrm{12.5}}/N_{\mathrm{\infty}}=0.79 \pm 0.02$}, in agreement with the values $0.71 \pm 0.12$ and $0.75 \pm 0.13$ measured for co- and cross-polarized interferometer arms, respectively. Overall, these results confirm the good correspondence between our measurements and our theoretical description of the interferometer.

In order to calculate the expected two-photon interference visibility $v$ for cavity-coupled NTDs, we solve Eqns.~\ref{Eq:cavitypop1}~--~\ref{Eq:cavitypop2} for the cavity population $\rho_{c}$ and evaluate ~\cite{Grange2015}:
\begin{equation}
    v = \frac{\int_0^{\infty} d t \rho_{\mathrm{c}}^2(t) \int_0^{\infty} d \tau e^{-\Gamma_{\mathrm{c}} \tau}}{\frac{1}{2}\left|\int_0^{\infty} d t \rho_{\mathrm{c}}(t)\right|^2}.
\end{equation}
In this equation, $\Gamma_c = \kappa + R$, with $\Gamma_c \approx \kappa$ in our system. With this expression, we find $v = 0.3$ for the theoretically expected visibility without significant dependence on $R$ within the previously estimated range ($0.13 - 2.0~\mu$eV).

Finally, to estimate the two-photon interference $v$ for free space NTDs, we set the coupling rate in Eqns.~\ref{Eq:cavitypop1}~--~\ref{Eq:cavitypop2} to zero, $R=0$, to obtain the free-space bright state decay as:
\begin{equation}
    \rho_{\mathrm{b}} (t) = \rho_b(0)\left(A_{fast} e^{-t/\tau_{\mathrm{fast}}} + A_{slow} e^{-t/\tau_{\mathrm{slow}}}\right). 
    \label{Eq:BiexponentialDecayMethods}
\end{equation} 
In this expression, $A_{fast/slow}\in{0, 1}$ are amplitude factors which obey ${A_{fast}+A_{slow}=1}$. We assume that fast and slow process have visibilites ${v_{\mathrm{fast/slow}} = T_2/(2\tau_{\mathrm{fast/slow}})}$, where the coherence time is given by the dephasing time as $T_2\approx1/\gamma^* = 80$~fs. All simulation results presented in the main text were obtained for a decay with free-space parameters $A_{fast}=0.92$ and $A_{slow} = 0.08$.

The fraction of the population which decays via the fast and slow process, respectively, is quantified by the relative fractional amplitudes $\tilde{A}_{\mathrm{fast/slow}}$ given by \cite{Gokus2010}:
\begin{equation}
    \tilde{A}_{\mathrm{fast/slow}} = \frac{A_{\mathrm{fast/slow}}\tau_{\mathrm{fast/slow}}}{A_{fast} \tau_{\mathrm{fast}} + A_{slow} \tau_{\mathrm{slow}}}.
    \label{Eq:Atilde}
\end{equation}
For the emitter dynamics considered in our simulation, we find $\tilde{A}_{fast}=0.34$ and $\tilde{A}_{slow} = 0.66$. We estimate $v$ as a weighted sum of visibilities for fast and slow process, $v = \tilde{A}_{\mathrm{fast}}v_{\mathrm{fast}} + \tilde{A}_{\mathrm{slow}}v_{\mathrm{slow}}$ to arrive at a vanishingly small free-space visibility $v= 0.003$.

\section*{Supplementary Note 5 - Two-photon interference timescale} 
\label{sec:TPITime}
\begin{figure}[t]
    \centering
    \includegraphics[scale=1]{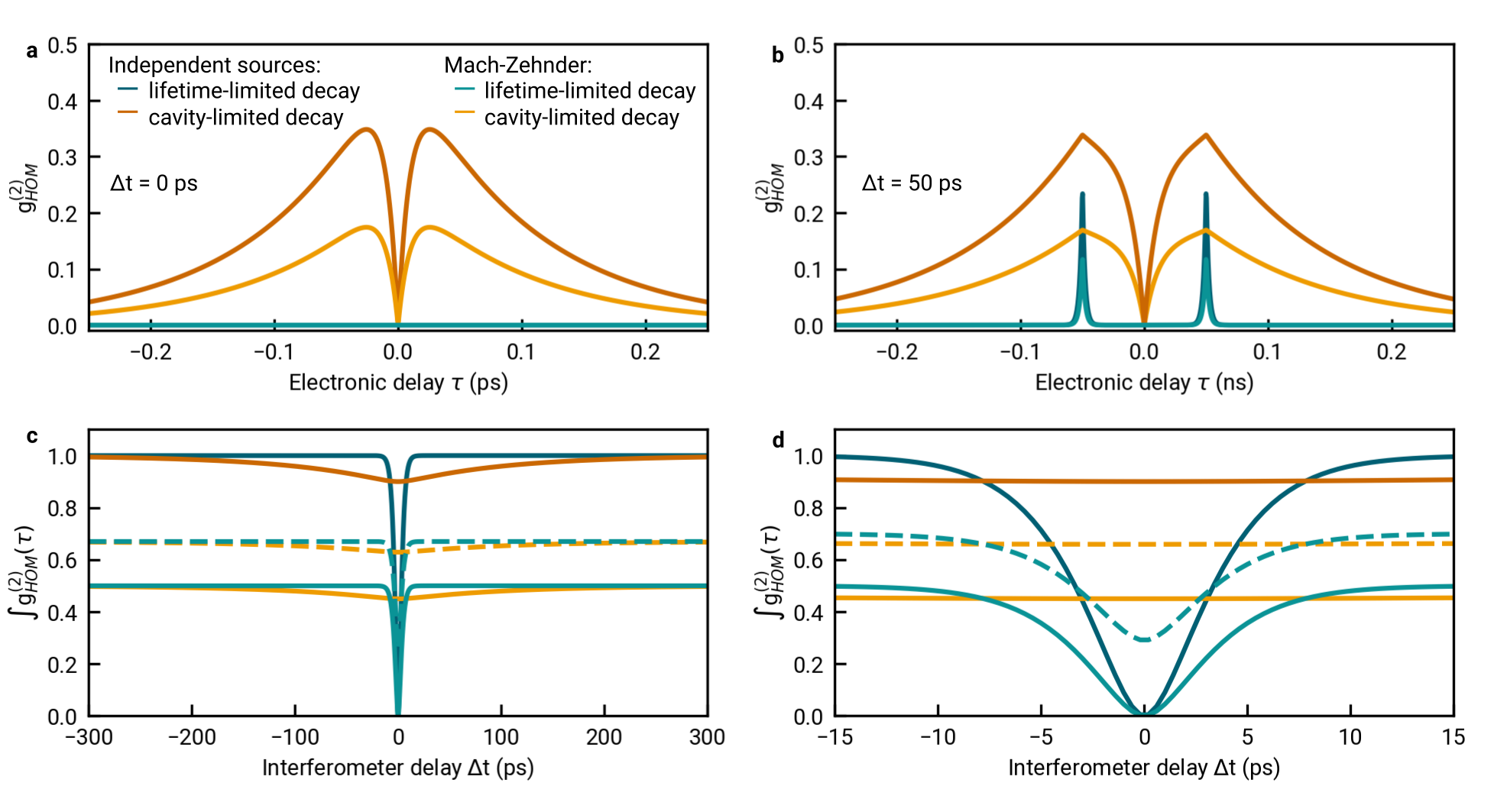}
    \caption{\textbf{Two-photon interference timescale.} \textbf{a}, \textbf{b}, HOM autocorrelation function as a function of electronic delay $\tau$, for interferometer delay $\Delta t = 0$~ps (\textbf{a}) and $\Delta t = 50$~ps (\textbf{b}). \textbf{c}, \textbf{d}, Integrated HOM autocorrelation function as a function of interferometer delay $\Delta t$. In all panels, the limiting cases of a fast, lifetime limited decay with lifetime 2~ps, and a slow decay with lifetime 100~ps and cavity-limited coherence time 20~ps are considered. Scenarios shown: two independent sources with unity single photon purity probed on a beamsplitter (dark green and orange solid lines); a single source with unity single photon purity probed in a Mach-Zehnder interferometer with 50:50 beamsplitters (blue and yellow solid lines); a single source with non-unity single photon purity (NTD~1) probed in our experimental Mach-Zehnder interferometer with imperfect beamsplitters (blue and orange dashed lines).}
    \label{figS3}
\end{figure}

As discussed in the main text, we associate the timescale $\tau_{\mathrm{HOM}}$ in the fit to the HOM dip in Fig.~4d with the emitter population lifetime. To explain this, we consider a monoexponentially decaying emitter with lifetime $T_1$ and coherence time $T_2$. Each data point in Fig.~4d is obtained from a correlation histogram as in Fig.~4c of the main text. When probing two independent emitters with unity single photon purity on a beamsplitter, the peak around time delay $\tau = 0$ in such a histogram is described by~\cite{Bylander2003}:
\begin{equation}
    g^{(2)}_{\mathrm{HOM}}(\tau) = \frac{1}{4} e^{-\left|\tau-\Delta t\right|/T_1} + \frac{1}{4} e^{-\left|\tau+\Delta t\right|/T_1} 
     - \frac{1}{2} e^{-\left|\tau\right|(2/T_2-1/T_1)-\left|\tau-\Delta t\right|/(2T_1)-\left|\tau+\Delta t\right|/(2T_1)}.
     \label{Eq:Bylander1}
\end{equation}
When probing a single emitter in a Mach-Zehnder interferometer with 50:50 beamsplitters, a prefactor of 1/2 has to be included to account for reduced coincidence probability around $\tau = 0$~\cite{Loredo2016, Ollivier2021}. 

We now consider the limiting cases of a lifetime-limited fast decay ($T_1=2$~ps, $T_2=4$~ps) and a cavity-limited slow decay ($T_1=100$~ps, $T_2=20$~ps given by the measured cavity lifetime), featured by two independent emitters and a single emitter probed in a Mach-Zehnder interferometer. We plot the results of Eq.~\ref{Eq:Bylander1} for each of these cases in Supplementary~Fig.~3a and b, with interferometer delay $\Delta t = 0$~ps and $\Delta t = 50$~ps, respectively. The lifetime-limited coherence of the fast decay enables two-photon interference, which results in vanishing correlation counts. 

As obvious from Eq.~\ref{Eq:Bylander1} and Supplementary~Fig.~3a and b, the coherence time $T_2$ can in principle be probed by varying the electronic delay $\tau$. By contrast, tuning of the interferometer delay changes the photon arrival time at the beamsplitter, such that this measurement probes the emitter population lifetime. In our experiment, the histogram bin size is much larger than the population lifetime and coherence time, and thus the histogram peak at $\tau = 0$ is given by~\cite{Bylander2003}:
\begin{eqnarray}
    g^{(2)}_{\mathrm{HOM}}(\tau=0) = \int_{N_0} g^{(2)}_{\mathrm{HOM}}(\tau) d\tau = \frac{1}{2}\left[1-\frac{T_2}{2T_1} e^{-2\left|\Delta t\right|/T_2} -\frac{1}{2T_1/T_2-1}\left(e^{-\left|\Delta t\right|/T_1} - e^{-2\left|\Delta t\right|/T_2} \right)\right],
    \label{Eq:Bylander2}
\end{eqnarray}
where integration is carried out over all counts in the histogram bin at $\tau = 0$ (c.f. Supplementary~Fig.~1). In Supplementary~Fig.~3c and d, we plot Eq.~\ref{Eq:Bylander2} for the limiting cases considered above. In addition, we also include the case of NTD~1 (as a single source with non-unity single photon purity) probed in our experimental Mach-Zehnder interferometer with imperfect beamsplitters described in Supplementary~Note~4. As explained above, in this case the values of $g^{(2)}_{\mathrm{HOM}}(\tau=0)$ are offset due to interferometer imbalance and nonzero $g^{(2)}_{\mathrm{HBT}}(0)$.

As obvious from Supplementary~Fig.~3c and d, the slow decay process will have an associated HOM timescale of 100~ps at two-photon interference visibility of around 0.1. By contrast, the fast decay process will have an associated HOM timescale of 2~ps, and unity interference visibility (HOM correlations do not vanish at $\Delta t = 0$~ps due to nonzero $g^{(2)}_{\mathrm{HBT}}(0)$). This motivates the interpretation presented in the main text: the fast decay process results in photons with near-unity visibility, which is reduced to our measured value of around 0.6 by photons generated at reduced indistinguishability via the slow process in the PL decay of Fig.~4e of the main text.   

Finally, we note that for highly indistinguishable photons with $T_2 \approx 2 T_1$, Eq.~\ref{Eq:Bylander2} simplifies to:

\begin{eqnarray}
    g^{(2)}_{\mathrm{HOM}}(\tau=0) \approx \frac{1}{2}\left(1-\frac{T_2}{2T_1} e^{-\left|\Delta t\right|/T_1}  \right).
    \label{Eq:Bylander3}
\end{eqnarray}
The function used to fit the data in Fig.~4d of the main text has the same functional form.

\begin{figure}[b!]
    \centering
    \includegraphics[scale=1]{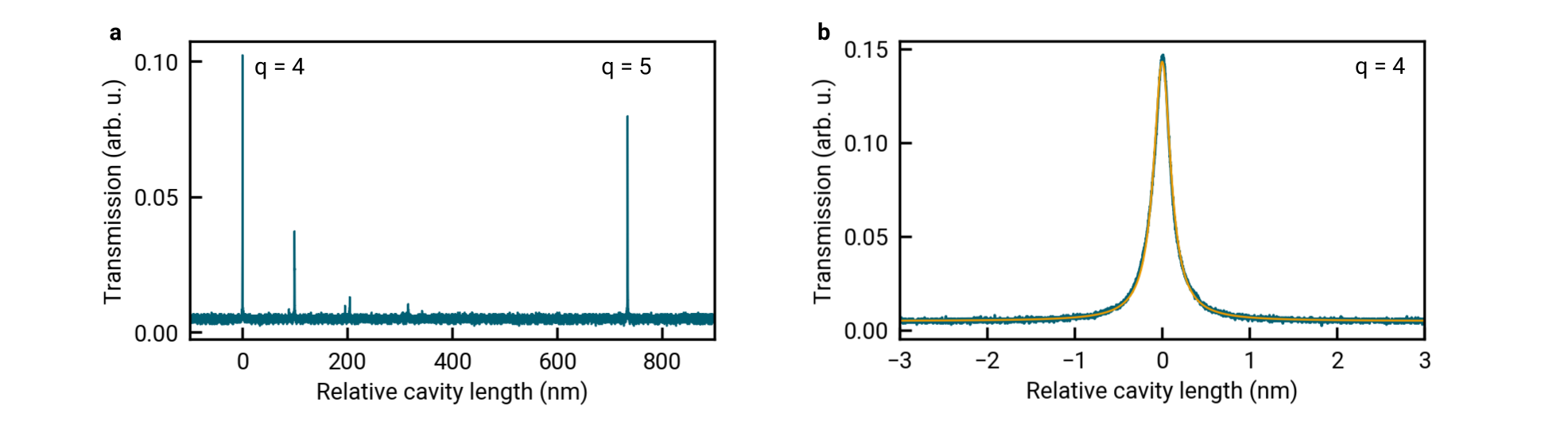}
    \caption{\textbf{Experimental cavity linewidth.} \textbf{a}, Cavity transmission at a wavelength of 1468.2~nm as a function of cavity length, which is tuned over 1.3 free spectral ranges \textbf{b}, Close-up on the resonance corresponding to longitudinal mode order $q=4$ shown in \textbf{a} and best fit of a Lorentzian line profile (solid orange line) with a linewidth of $\kappa = 34.8\ \mu$eV.}
    \label{figS4}
\end{figure}

\section*{Supplementary Note 6 - Experimental cavity linewidth} 
\label{sec:KappaMeas}

The cavity linewidth $\kappa$ was obtained from the cavity transmission of a diode laser with a wavelength measured as 1468.2~nm and limited by the resolution of the spectrometer. Supplementary~Fig.~4a shows the measured transmission as the cavity length is tuned by 1.3 free spectral ranges (FSRs), obtained after laterally positioning the cavity mode on the bare mirror away from NTDs. The two resonances at maximum transmission correspond to the $\mathrm{TEM}_{00}$-modes of the lowest accessible longitudinal mode orders $q=4$ and 5, respectively. Their distance corresponds to exactly one FSR, or half the wavelength, and was used to compute the time-dependent change in cavity length. The additional resonances stem from higher order TEM modes. From Lorentzian fits to the transmission of the $q=4$ resonance as in Supplementary~Fig.~4b, we obtained an averaged cavity linewidth $\kappa = 35.4 \pm 0.1\ \mu$eV from ten repetitions of the measurement.